\documentclass[a4paper,11pt]{article}
\pdfoutput=1 

\usepackage{jheppub} 
\usepackage[T1]{fontenc} 

\title{Differential equations, recurrence relations, and quadratic constraints for $L$-loop two-point massive tadpoles and propagators.}

\author{Roman N. Lee}
\author{and Andrei A. Pomeransky}
\affiliation{The Budker Institute of Nuclear Physics, 630090, Novosibirsk}
\emailAdd{r.n.lee@inp.nsk.su}
\emailAdd{a.a.pomeransky@inp.nsk.su}

\abstract{
We consider $L$-loop two-point tadpole (watermelon) integral with arbitrary masses, regularized both dimensionally and analytically. We derive differential equation system and recurrence relations (shifts of dimension and denominator powers). 
Since the $L$-loop sunrise integral corresponds to the $(L+1)$-loop watermelon integral with one cut line, our results are equally applicable to the former.
The obtained differential system has a Pfaffian form and is linear in dimension and analytic regularization parameters. In general case, the solutions of this system can be expressed in terms of the Lauricella functions $F_C^{(L)}$ with generic parameters. Therefore, as a by-product, we obtain, to our knowledge for the first time, the Pfaffian system for $F_C^{(L)}$  for arbitrary $L$. The obtained system has no apparent singularities. Near odd dimension and integer denominator powers the system can be easily transformed into canonical form. 
Using the symmetry properties of the matrix in the right-hand side of the differential system, we obtain quadratic constraints for the expansion of solutions near integer dimension and denominator powers. In particular, we obtain quadratic constraints for Bessel moments similar to those discovered by Broadhurst and Roberts.

}

\newcommand{\e}{\epsilon}
\newcommand{\mub}{\boldsymbol{\mu}}
\newcommand{\mb}{\boldsymbol{m}}
\newcommand{\IKM}{\mathrm{IKM}}
\begin{document} 
\maketitle
\flushbottom

\section{Introduction}
\label{sec:intro}

Multiloop massive sunrise and watermelon integrals  (see Fig. \ref{fig:tadpole}) are ubiquitous in various physical applications. Being seemingly simple, they, in fact, bring with themselves a lot of complexity in any calculation. The reason is that their $\e$-expansion can not be expressed in terms of the standard functional basis, viz., multiple polylogarithms \cite{Goncharov1998,Weinzierl:2002hv}. Already starting from two loops one observes complicated iterated integrals involving modular forms and/or elliptic curves \cite{Adams2017b,Broedel2018c}. For $L$ loops, the $\e$ expansion of massive sunrise integrals is likely to involve periods of certain $(L-1)$-dimensional Calabi-Yau varieties (cf. Refs. \cite{arXiv0407327v1.math,vanhove2019feynman}). 

For the general case of different masses the closed expressions for these integrals  are known since Ref. \cite{Berends:1993ee} (see also Ref. \cite{Kalmykov:2016lxx}) in terms of Lauricella functions $F_C^{(L)}$ defined via multiple hypergeometric series. Although these functions have a long history of investigation, starting from \cite{lauricella1893sulle}, and much is known about their properties, still many questions remain unanswered. In particular, the functions which appear in coefficients of their expansion in the Laurent series in parameters are not well investigated. Meanwhile, these coefficients are the most important from the point of view of multiloop calculations. From the purely mathematical point of view, among the four flavors of Lauricella functions $F_A$, $F_B$, $F_C$, and  $F_D$, the functions $F_C^{(L)}$ are the only ones for which the differential system in Pfaffian form is not yet known for $L>2$. 

In the present paper we derive the Pfaffian system of the differential equations for the $L$-loop watermelon and sunrise graphs with arbitrary masses regularized both dimensionally and analytically. Similar to Refs. \cite{Berends:1993ee,Kalmykov:2016lxx}, we find the fundamental system of solutions in terms of the Lauricella functions $F_C^{(L)}$ with generic parameters and arguments. In this way, our results give a Pfaffian system for Lauricella $F_C^{(L)}$ for all $L$, to the best of our knowledge, for the first time\footnote{For the case $L=2$ a Pfaffian system was found in Ref. \cite{kato1988pfaffian}.}. 

In addition to the differential system, we also derive the operators shifting dimension or the powers of denominators. From the point of view of $F_C^{(L)}$ functions these operators provide a certain form of the contiguity relations. Near $D=1$  we reduce the differential system to canonical form which allows one to obtain the coefficients of expansion in dimension and analytic regulators in terms of the Goncharov's polylogarithms.

Last but not least, we observe that the matrix in the right-hand side of the obtained differential system has special features which result in the existence of the bilinear relations for the solutions of the systems that differ in the sign of the right-hand side (cf. Ref. \cite{Goto2013}).
Combined with the shift operators, these relations allow us to obtain the infinite set of the quadratic relations for the coefficients of expansion in dimension and analytic regulators near any integer point, including the most interesting case of even $D$.

\section{Differential equations in Pfaffian form}
\begin{figure}\centering
	\includegraphics{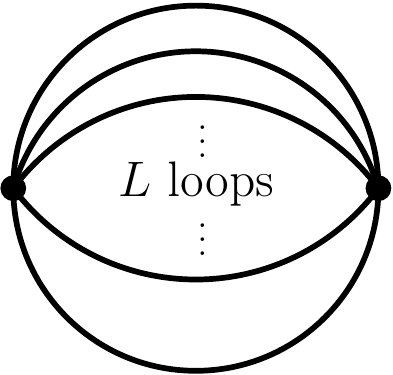}
	\caption{$L$-loop two-point massive tadpole. Each line corresponds to the propagator with arbitrary mass and power.}
	\label{fig:tadpole}
\end{figure}

We consider the $L$-loop watermelon integral, Fig.\ref{fig:tadpole}, with different masses and powers of denominators (exponents),

\begin{equation}\label{eq:tadpolePE}
\mathcal{T}(m_0,\ldots,m_L)=
2^{D-1}i\pi^{D/2}\Gamma\left(\tfrac{D}{2}\right)
\int\delta\left(\textstyle\sum_{l=0}^{L}p_l\right)\prod_{l=0}^{L}\left[\frac{d^{D}p_{l}}{i\pi^{D/2}}\frac{m_l\Gamma\left(\alpha_{l}\right)}{\left(m_{l}^{2}-p_{l}^{2}-i0\right)^{\alpha_{l}}}\right]\,.
\end{equation}
Here we imply the Lorentzian signature, so that $p^2=p_0^2-\boldsymbol{p}^2$. Note that we have chosen a somewhat unconventional overall normalization factor $2^{D-1}\Gamma\left(\tfrac{D}{2}\right)\prod_l m_l\Gamma\left(\alpha_{l}\right)$ in order to simplify some expressions below.

Before moving any further let us discuss the applicability of our present consideration to the sunrise integral
\begin{equation}\label{eq:sunrisePE}
\mathcal{S}(m_0,\ldots,m_L|q^2)=
2^{D-1}i\pi^{D/2}\Gamma\left(\tfrac{D}{2}\right)
\int
\delta\left(\textstyle q+\sum_{l=0}^{L}p_k\right)\prod_{l=0}^{L}\left[\frac{d^{D}p_{l}}{i\pi^{D/2}}\frac{m_k\Gamma\left(\alpha_{l}\right)}{\left(m_{l}^{2}-p_{l}^{2}-i0\right)^{\alpha_{l}}}\right].
\end{equation}
The obvious relation between $\mathcal{S}$ and  $\mathcal{T}$ is 
$ \mathcal{S}|_{q^2=0} = \mathcal{T}$. However, we can also represent the $(L-1)$-loop sunrise integral as an $L$-loop watermelon integral with one cut line:
\begin{multline}\label{eq:sunrisePE1}
(m_0)^{D-1}\mathcal{S}(m_1,\ldots,m_L|m_0^2)=
2^{D-2}i\pi^{D/2-1}\Gamma^2\left(\tfrac{D}{2}\right)\\
\times
\int
\delta\left(\textstyle \sum_{l=0}^{L+1}p_k\right)
\frac{d^{D}p_{0}}{i\pi^{D/2}}\left[\frac{m_{0}}{m_{0}^{2}-p_{0}^{2}-i0}-\frac{m_{0}}{m_{0}^{2}-p_{0}^{2}+i0}\right]\\
\times\prod_{l=1}^{L}\left[\frac{d^{D}p_{l}}{i\pi^{D/2}}\frac{m_k\Gamma\left(\alpha_{l}\right)}{\left(m_{l}^{2}-p_{l}^{2}-i0\right)^{\alpha_{l}}}\right]\,,
\end{multline}
where $m_{0}=\sqrt{q^2}$. The two terms in square brackets on the last line correspond to two different deformations of the integration contour over $p_0^0$. As the differential system we are looking for is not aware of the integration contours, it is clear that $m_{0}^{D-1}\mathcal{S}(m_1,\ldots,m_L|m_0^2)$ satisfies the same differential equations as  $\mathcal{T}(m_0,\ldots,m_L)$ with $\alpha_0=1$. Therefore, in the rest of the paper we will concentrate on the watermelon integral $\mathcal{T}$ only.

After the Wick rotation we can rewrite Eq. \eqref{eq:tadpolePE} as
\begin{equation}\label{eq:tadpole}
\mathcal{T}=
2^{D-1}\pi^{D/2}\Gamma\left(\tfrac{D}{2}\right)
\int
\delta\left(\textstyle\sum_{l=0}^{L}p_k\right)\prod_{l=0}^{L}\left[\frac{d^{D}p_{l}}{\pi^{D/2}}\frac{m_k\Gamma\left(\alpha_{l}\right)}{\left(p_{l}^{2}+m_{l}^{2}\right)^{\alpha_{l}}}\right]\,,
\end{equation}
where now $p^2=p_0^2+\boldsymbol{p}^2$. Representing the function $\delta(p)$ as $\intop \frac{d^{D}x}{(2\pi)^D}e^{ipx} $, we obtain
\begin{equation}\label{eq:tadpole-xspace}
\mathcal{T}
=\intop_0^{\infty} d x\,x^{D-1} \prod_{l=0}^{L} \int\frac{d^{D}p}{\pi^{D/2}}\frac{m_k\Gamma(\alpha_k)e^{ipx}}{(p^{2}+m_k^{2})^{\alpha_k}}
=\intop_0^{\infty} d x\,x^{D-1} \prod_{l=0}^{L}P_0(\mu_k,m_k,x)\,.
\end{equation}

As we shall see immediately, the integral $\int\frac{d^{D}p}{\pi^{D/2}}\frac{m\Gamma(\alpha)e^{ipx}}{(p^{2}+m^2)^{\alpha}}$ depends on $D$ and $\alpha$ only via the combination $\mu=D-2\alpha+1$ and we have, therefore, introduced
\begin{multline}\label{eq:P0}
	 P_0(\mu,m,x) =m\Gamma(\alpha)\int\frac{d^{D}p}{\pi^{D/2}}\frac{e^{ipx}}{(p^{2}+m^{2})^{\alpha}}
	 =m\intop_0^{\infty} d\lambda\lambda^{\alpha-1}\int\frac{d^{D}p}{\pi^{D/2}}e^{ipx-\lambda\left(p^{2}+m^{2}\right)}\\
	 =m\intop_0^{\infty} d\lambda\lambda^{\frac{1-\mu}2-1}e^{-\lambda m^{2}-\frac{x^{2}}{4\lambda}}=2m\left(\frac{x}{2m}\right)^{\frac{1-\mu}2}K_{\frac{\mu-1}2}\left(mx\right)\,.
\end{multline}
Here $K_\nu(y)$ is the Macdonald function. 

In order to derive the differential equations for this integral, we introduce the set of $2^{L+1}$ functions labeled by a binary number (a string of zeros and ones) $\boldsymbol{a}=a_0a_1\ldots a_L$,
\begin{equation}
	T_{\boldsymbol{a}}\left(D, \mub,\mb\right)=\intop_0^{\infty} d x\,x^{D-1}t_{\boldsymbol{a}}\,,
	\label{eq:T_a}
\end{equation} 
where 
\begin{equation}
		t_{\boldsymbol{a}}=\prod_{l=0}^{L}P_{a_l}(\mu_l,m_l,x)\,,
		\label{eq:t_a}
\end{equation}
$P_0$ is defined in eq. \eqref{eq:P0}, and 
\begin{multline}\label{eq:P1}
P_{1}(\mu,m,x) =\frac{1}{m}\partial_{x}P_0(\mu,m,x) =-\frac{x}{2}\intop_0^{\infty} d\lambda\lambda^{-\frac{1+\mu}2-1}e^{-\lambda m^{2}-\frac{x^{2}}{4\lambda}}\\
 =
-2m\left(\frac{x}{2m}\right)^{\frac{1-\mu}2}K_{\frac{\mu+1}2}\left(mx\right)\,.
\end{multline}
The original integral \eqref{eq:tadpole} is recovered as $\mathcal{T}=T_{\boldsymbol{0}}\left(D, \mub,\mb\right)$. The rationale behind introducing auxiliary $2^{L+1}-1$ functions is that it turns out to be possible to derive a closed system of linear differential equations for $T_{\boldsymbol{a}}$. Note that all components of $T_{\boldsymbol{a}}$ can be expressed via $\mathcal{T}$ with shifted dimension and indices:
\begin{multline}
	T_{\boldsymbol{a}}(D,\mub) = \prod_l\left(\frac{-1}{2m_l}\right)^{a_l} \mathcal{T}(D\to D+|\boldsymbol{a}|,\mu_l\to \mu_l+2a_l)\\
	 = \prod_l\left(\frac{-1}{2m_l}\right)^{a_l} \mathcal{T}(D\to D+|\boldsymbol{a}|,\alpha_l\to \alpha_l+|\boldsymbol{a}|/2-a_l)\,,
\end{multline}
where $|\boldsymbol{a}|=\sum_l a_l$.
It is convenient to treat the $2^{L+1}$ quantities $V_{\boldsymbol{a}}$ as components of the vector $V$ in $\underbrace{\mathbb{C}^2\otimes\ldots \otimes \mathbb{C}^2}_{L+1 \text{ factors}}$, where $V_{\boldsymbol{a}}$ can be either $T_{\boldsymbol{a}}$ or  $t_{\boldsymbol{a}}$. In what follows, we assume that the factors in $\mathbb{C}^2\otimes\ldots \otimes \mathbb{C}^2$ are numbered starting from zero, so that, e.g., $L$-th factor is the right-most one. Besides, to lighten notation we often omit the arguments of functions.  

In order to derive the differential system for $T_{\boldsymbol{a}}$ we differentiate $P_{0,1}(\mu,m,x)$ with respect to $m$ and use the differential equation for $K_\nu(y)$. We obtain 
\begin{align}
	m\partial_{m}P_0	&=mxP_1+\mu P_0\,,\nonumber\\
	m\partial_{m}P_1	&=mxP_0\,.
\end{align}
In matrix notations we may rewrite this as
\begin{equation}\label{eq:mdmP}
	m\partial_{m}P = (mx\sigma + \mu \bar{n})P\,,
\end{equation}
where $P=\begin{pmatrix}P_0\\ P_1\end{pmatrix}$, 
$\sigma=\sigma_x=\begin{pmatrix} 0& 1 \\ 1&0 \end{pmatrix}$, and $\bar{n}=1-n=\frac12(1+\sigma_z)=\begin{pmatrix} 1& 0 \\ 0&0 \end{pmatrix}$.

Using these formulae, we see that the expression for the derivative of $T$, in addition to $T$ multiplied by a matrix, contains the integrals of the form $\intop_0^{\infty} d x\,x^{D}\, t$.
Note the additional factor $x$ in the integrand as compared to the definition \eqref{eq:T_a}. 
In order to get rid of this factor, we consider the identity $\int dx \partial_x (x^D t)=0$. Explicitly differentiating and using the relation
\begin{equation}
x\partial_{x} P = (m\partial_m-\mu)P = (mx\sigma - \mu n)P\,,
\end{equation}
which, again, follows from the differential equation for $K_{\nu}$, we obtain
\begin{equation}
\intop_0^{\infty} d x\,x^{D-1} \left(xM-W\right)t =0\,.
\end{equation}
Here
\begin{equation}
	M=\sum_{l=0}^L m_{l}\sigma_{l}\,,\quad W=\sum_{l=0}^L\mu_{l}n_{l}-D\,,
\end{equation}
and the operators $\sigma_l$ and $n_l$ act in $l$-th factor of $\mathbb{C}^2\otimes\ldots \otimes \mathbb{C}^2$ as $\sigma$ and $n$, respectively. Therefore,
\begin{equation}\label{eq:xaction}
\intop_0^{\infty} d x\,x^{D}\, t = M^{-1}W\, T
\end{equation}
whenever $M^{-1}$ exists. Since the operators $\sigma_l$ commute with each other and their eigenvalues are $\eta_l=\pm 1$, the eigenvalues of $M$ are $\sum_l \eta_l m_l$. Therefore, $M$ is not invertible iff 
\begin{equation}\label{eq:singular}
	\sum_l \eta_l m_l = 0
\end{equation} for some choice of signs  $\eta_l$.

Thus, using Eqs. \eqref{eq:mdmP} and  \eqref{eq:xaction} we obtain the following expression for the derivative of $T$ with respect to $m_k$
\begin{equation}\label{eq:dm_k}
	\partial_{m_{l}}T=A_{l}T,\qquad A_{l}=\sigma_{l}M^{-1}W + \frac{\mu_{l}\bar{n}_{l}}{m_{l}}\,.
\end{equation}
Equivalently, the above equations can be written as 
\begin{gather}\label{eq:dlog}
	dT=A\,T\,,\\
	A = \sum_{l=0}^{L} A_l dm_l =dM\,M^{-1}W + \sum_{l=0}^{L} \mu_{l}\bar{n}_{l}\frac{dm_l}{m_l} \,. 
\end{gather}
Remarkably, the differential form $A$ is closed ($dA=0$) which can be easily proved if one takes into account the pairwise commutativity of $\sigma_l$. Then the integrability condition $dA=A\wedge A$ requires that $A\wedge A=0$, see Appendix \ref{sec:Compatibility} for the derivation. The system \eqref{eq:dlog} equipped with the condition $dA=0$ is said to be in Pfaffian form. We observe that all singularities of the differential form $A$ are located on the $L+1$ hyperplanes defined by $m_l=0$ and $2^L$ hyperplanes defined by Eq. \eqref{eq:singular} (cf. Ref. \cite{Goto2013}). The latter hyperplanes correspond to vanishing of the eigenvalues of the matrix $M$.

Let us note that the differential system \eqref{eq:dlog} in fact splits into two separate systems, each for $2^L$ functions $T_{\boldsymbol{a}}$ with $|\boldsymbol{a}|$ being either even or odd. This due to the commutativity of $A$ with the parity operator 
\begin{equation}\label{eq:Parity}
	\mathcal{P}=\left(-1\right)^{\sum_i n_i}=\prod_i (1-2n_i)\,.
\end{equation}
Of course, we are mostly interested in the subsystem for even components which involves our original tadpole integral $\mathcal{T}$.

\section{Shifting exponents and dimension}
Let us now obtain the recurrence relations in $\alpha_k$ and $D$. For the former we use the identities
\begin{align}
	P_0\left(\mu-2,m,x\right)	&=m \intop_0^{\infty} d\lambda\lambda^{\frac{1-\mu}2}e^{-\lambda m^{2}-\frac{x^{2}}{4\lambda}}=-\frac{1}{2m^2}(m\partial_{m}-1)P_0\left(\mu,m,x\right)\,,\\
	P_1\left(\mu-2,m,x\right)&=-\frac{x}{2}\intop_0^{\infty} d\lambda\lambda^{-\frac{1+\mu}2}e^{-\lambda m^{2}-\frac{x^{2}}{4\lambda}}=-\frac{1}{2m}\partial_m P_1\left(\mu,m,x\right)\,.
\end{align}
Then, using Eq. \eqref{eq:dm_k}, we have
\begin{equation}\label{eq:alphashift}
	T(\mub-2\boldsymbol{e}_l) = R_{l}\left(D,\mub\right) T(\mub)\,,
\end{equation}
where
\begin{equation}
	R_{l}\left(D,\mub\right)=-\frac{1}{2m_{l}}\left(A_{l}-\frac{\bar{n}_{l}}{m_{l}}\right)
	=-\frac{1}{2m_{l}}\left(\sigma_{l}M^{-1}W+\frac{\left(\mu_{l}-1\right)\bar{n}_{l}}{m_{l}}\right)
\end{equation}
and $\boldsymbol{e}_l =(\ldots,0,\underset{l\text{-th}}{1},0,\ldots)$ is the vector with $l$-th component equal to $1$ and all other components equal to zero. 

Let us now derive the operator which shifts the dimension. Note that the identity \eqref{eq:xaction} defines the operator 
\begin{equation}
	R(D,\mub)=M^{-1} W\,,
\end{equation} 
which shifts $D$ by $+1$ at fixed $\mub$, i.e.
\begin{equation}\label{eq:Dshift}
	T(D+1,\mub)=R(D,\mub) T(D,\mub)\,.
\end{equation}
In order to obtain the operator which shifts $D$ at fixed $\boldsymbol{\alpha}$, we should first use the operators $R_l$ to shift $\mu\to \mu-2$ and then shift the dimension by $-2$ at fixed $\mu$ using the operator $R^{-1}$. We obtain
\begin{equation}\label{eq:lDRR}
T\left(D-2,\mub-2\boldsymbol{e}\right)=\mathfrak{R}(D,\mub)T(D,\mub)\,,
\end{equation}
where 
\begin{equation}\label{eq:Dmushift}
\mathfrak{R}(D,\mub)=
R^{-1}(D-2,\mub-2\boldsymbol{e})
R^{-1}(D-1,\mub-2\boldsymbol{e})
\prod_{l=0}^L R_l\left(D,\mub-2\textstyle\sum_{j=l+1}^{L}\boldsymbol{e}_j\right)
\end{equation}
is the operator which shifts $D$ by $-2$ at fixed $\boldsymbol{\alpha}$. Here and below $\boldsymbol{e}=\sum_{j=0}^{L}\boldsymbol{e}_j=(1,1,\ldots,1)$ and the product of matrices corresponds to the multiplication from the left to the right, so that $\prod_{l=0}^{L} R_l = R_0\ldots R_L$.

\section{Basis of solution space}\label{sec:basis}
While the function $T$ as defined in Eq. \eqref{eq:T_a} is a specific solution of the differential system \eqref{eq:dlog} and recurrence relations \eqref{eq:alphashift} and \eqref{eq:Dshift}, it is natural to consider the linear space of all solutions of these equations. In the present section we fix a basis of $2^{L+1}$ functions in this space. In order to do this, we note that the derivation of the differential equations and recurrence relations in Sections 2 and 3 can be explicitly repeated if some of the Macdonald functions $K_\nu$ are replaced with $\frac{I_{\pm\nu}}{\sin\pi\nu}$. More specifically, let us make the following replacement in the definition of $t_{\boldsymbol{a}}$, Eq. \eqref{eq:t_a}: 
\begin{equation}
	K_{\frac{\mu_l\pm1}2}\left(m_lx\right) \to \mp\frac{\sigma_l \pi }{2\cos\frac{\pi\mu_l}2}I_{\sigma_l\frac{\mu_l\pm1}2}\left(m_lx\right)\,,
\end{equation}
for all $l\in S$, where $S$ is some subset of $E = \{0,1,\ldots,L\}$. Here each $\sigma_l$ can be either $+1$ or $-1$. Defined in this way, the function $T_{\boldsymbol{a}}$ formally satisfies the differential and difference equations \eqref{eq:dlog}, \eqref{eq:alphashift}, and \eqref{eq:Dshift}. There may be however a problem with the convergence of the integral over $x$ due to the exponential growth of the modified Bessel functions. For all masses being positive, the convergence condition is $\sum_{l\in S} m_l<\sum_{l\in E\backslash S} m_l$. In particular, $E\backslash S$ should not be empty. Therefore, we take $E\backslash S=\{0\}$, i.e.  $S=\{1,\ldots, L\},\,$ and define the set of functions
\begin{gather}\label{eq:Vrho}
	V^{(\boldsymbol{\rho})}=\intop_0^{\infty} dx\,x^{D-1} P(\mu_0,m_0,x)\otimes\bigotimes_{l=1}^L Q^{(\rho_l)}(\mu_l,m_l,x)\,,\\
	\label{eq:Qdef}
	Q^{(\rho)}(\mu,m,x) = {Q^{(\rho)}_0(\mu,m,x) \choose Q^{(\rho)}_1(\mu,m,x)} = \frac{\pi(-1)^\rho}{2\cos\frac{\pi\mu}2}2m\left(\frac{x}{2m}\right)^{\frac{1-\mu}2}\begin{pmatrix}I_{(-1)^\rho\frac{\mu-1}{2}}\left(mx\right)\\
	I_{(-1)^\rho\frac{\mu+1}{2}}\left(mx\right)\end{pmatrix}
	\,,
\end{gather}
where $\boldsymbol{\rho}=\rho_1\rho_2\ldots\rho_L$ is a binary number. 
We assume that $m_l>0$ and  $m_0>\sum_{l=1}^{L} m_l$, so that the integrand decays exponentially at large $x$. Using the identity
\begin{equation}
K_{\nu }(x)=\frac{\pi(I_{-\nu }(x)-I_{\nu }(x))}{2\sin \pi \nu}\,,
\end{equation}
we can express $T$, Eq. \eqref{eq:T_a}, via $V^{(\boldsymbol{\rho})}$ as
\begin{equation}\label{eq:TviaV}
	 T=\sum_{\boldsymbol{\rho}} V^{(\boldsymbol{\rho})}\,.
\end{equation}

The set of functions \eqref{eq:Vrho} contains  $2^L$ functions numbered by the vector $\boldsymbol{\rho}$. Meanwhile, the solution space of the differential system \eqref{eq:dlog} is $2^{L+1}$-dimensional. In order to obtain the whole set of solutions, we use the symmetry of the differential system related to the operator $\mathcal{P}$, Eq. \eqref{eq:Parity}. Therefore, we can define functions
\begin{equation}\label{eq:Vpm}
V^{(\varrho_0,\boldsymbol{\rho})}=\frac{1+(-1)^{\varrho_0+|\boldsymbol{\rho}|}\mathcal{P}}2 V^{(\boldsymbol{\rho})}\,,\quad \varrho_0=0,1,
\end{equation}
which are the solutions of the differential system. Note that we have introduced the additional factor $(-1)^{|\boldsymbol{\rho}|}$ for further convenience. In terms of the components we have
\begin{equation}\label{eq:Vpma}
	V^{(\varrho_0,\boldsymbol{\rho})}_{\boldsymbol{a}}=\frac{1+(-1)^{\varrho_0+|\boldsymbol{\rho}|+|\boldsymbol{a}|}}2 V^{(\boldsymbol{\rho})}_{\boldsymbol{a}}\,,
\end{equation}
so that the solutions with $\varrho_0 = |\boldsymbol{\rho}|\ (\mathrm{mod}\ 2)$ have nonzero components with even $|\boldsymbol{a}|$.

Together with the different choices of $\boldsymbol{\rho}$, this gives us $2^{L+1}$ solutions of the differential system \eqref{eq:dlog}. It is also easy to see that $\mathcal{P}R_l=R_l\mathcal{P}$ and, therefore, $V^{(\varrho_0,\boldsymbol{\rho})}$ is also a solution of the $\mu$-shifting recurrences \eqref{eq:alphashift}. As to the the operator $R$, Eq. \eqref{eq:Dshift}, we have $\mathcal{P}R=-R\mathcal{P}$ and therefore
\begin{equation}
	V^{(\varrho_0,\boldsymbol{\rho})}(D+1) = R(D)V^{(\bar\varrho_0,\boldsymbol{\rho})}(D)\,,
\end{equation}
where we used the notation $\bar{x}=1-x$.
Thus, the functions \eqref{eq:Vpm} satisfy the relations shifting $D$ by $2$,
\begin{equation}
	V^{(\varrho_0,\boldsymbol{\rho})}(D+2) = R(D+1)R(D)\ V^{(\varrho_0,\boldsymbol{\rho})}(D)\,,
\end{equation}
which follow from  Eq. \eqref{eq:Dshift}\footnote{Note, that we can also construct $2^{L+1}$ functions which satisfy \eqref{eq:Dshift}, e.g., by taking $V^{(\boldsymbol{\rho})}=V^{(0\boldsymbol{\rho})}+V^{(1\boldsymbol{\rho})}$ and $(V^{(0\boldsymbol{\rho})}-V^{(1\boldsymbol{\rho})})g(D)$, where $g$ is any antiperiodic function of $D$, i.e., $g(D+1)=-g(D)$.}.

Note that our basis functions $V^{(\varrho_0,\boldsymbol{\rho})}$ can be expressed via Lauricella function $F_C^{(L)}$ (cf. \cite{Berends:1993ee,Kalmykov:2016lxx}). This function is defined via the hypergeometric series
\begin{equation}
	F_C^{(L)}(b_1,b_2;c_1,\ldots c_L;x_1,\ldots,x_L)=\sum_{k_1=0}^{\infty}\ldots \sum_{k_L=0}^{\infty}
	\frac{(b_1)_{\sum k_i}(b_2)_{\sum k_i} x_1^{k_1}\ldots x_L^{k_L}}{(c_1)_{k_1}\ldots (c_L)_{k_L}\,k_1!\ldots k_L!}\,,
\end{equation}
where $(a)_k=\Gamma(a+k)/\Gamma(a)$ is the Pochhammer symbol.
In order to do this we use the expansion of the modified Bessel function $I_\nu$:
\begin{equation}
	I_{\nu}(x)= \sum_{k=0}^{\infty}\frac{(x/2)^{\nu+2k}}{k! {\Gamma (\nu + k+1)}}\,.
\end{equation}
Substituting this expansion in Eq. \eqref{eq:Vrho} and taking the integral over $x$ with the help of the identity
\begin{equation}
	\intop_0^\infty dx\,x^{\beta-1}K_{\nu}\left(x\right)=2^{\beta-2}\Gamma\left[\tfrac{\beta+\nu}{2}\right]\Gamma\left[\tfrac{\beta-\nu}{2}\right]\,,
\end{equation}
we obtain
\begin{multline}\label{eq:Lauricella}
	V^{(\boldsymbol{\rho})}_{\boldsymbol{a}}=2^{D-1}\left(-1\right)^{a_{0}}m_{0}^{-D+\sum_{l=0}^{L}\mu_l}\Gamma\left[b_1\right]\Gamma\left[b_2\right]\prod_{l=1}^{L}\left(-1\right)^{a_{l}}\Gamma\left[1-c_l\right]\left(\frac{m_{l}}{m_0}\right)^{c_l+\tfrac{\mu_l-1}2}\\
	\times F_{C}^{\left(L\right)}\left(b_1,b_2;c_1,\ldots c_L;\frac{m_{1}^{2}}{m_{0}^{2}},\ldots, \frac{m_{L}^{2}}{m_{0}^{2}}\right)\,,
\end{multline}
where
\begin{align}
	c_l&=1+(-1)^{\rho_l}\frac{\mu_{l}-(-1)^{a_l}}{2}\,,\nonumber\\
	b_1&=\frac{D+a_{0}}{2}+\frac12\sum_{l=1}^{L}\left[c_l-\frac{\mu_l+1}{2}\right]\,,\nonumber\\
	b_2&=b_1-a_0+\frac{1-\mu_0}{2}\,.
\end{align}
The functions $V^{(\varrho_0,\boldsymbol{\rho})}$ can be obtained from Eq. \eqref{eq:Lauricella} with the help of Eq. \eqref{eq:Vpma}. 

Note that 
\begin{equation}
	V^{({1\ldots 1})}_{{0\ldots 0}}\propto m_{0}^{-L-D+\sum_{l=0}^{L}\mu_l} [\prod_{l=1}^{L}m_l] F_{C}(\tilde{b}_1,\tilde{b}_2;\tilde{c}_1,\ldots \tilde{c}_L;x_1,\ldots x_L)\,,
\end{equation}
where $x_l=(m_l/m_0)^2$ and the parameters
\begin{equation}
\tilde{c}_l=\frac12\left(3-\mu_l\right)\,,\quad
\tilde{b}_1=\frac{D}{2}+\frac12\sum_{l=1}^{L}\left[1-\mu_l\right]\,,\quad
\tilde{b}_2=\tilde{b}_1+\frac{1-\mu_0}{2}
\end{equation}
are generic for generic $\mub$ and $D$. 
Therefore, $V^{({1\ldots 1})}_{{0\ldots 0}}$ is proportional to $ F_{C}^{(L)}$ with generic parameters. Recall that the system \eqref{eq:dlog} splits into two subsystems, for even and odd components. Thus, we see that the subsystem of $2^L$ equations for even components gives a Pfaffian system for Lauricella $F_C^{(L)}$.

\section{Bilinear relations and expansion near integer $D$ and $\mub$.}

The fundamental matrix of the differential system \eqref{eq:dlog} can be written in the form of path-ordered exponent
\begin{equation}\label{eq:pexp}
U(D,\mub,\mb)=\mathrm{Pexp}\left[\int_{\mathcal C} A\right]\,,
\end{equation}
where ${\mathcal C}$ is some path in $L+1$-dimensional space of mass parameters starting from some fixed point $\mb_0$ and ending in $\mb$\footnote{Note that the matrix $A$ has singularities when either $\det M=0$ or $m_l=0$. Naturally, the path $\mathcal{C}$ should not intersect the singular locus of $A$, and, therefore, $U$ depends not only on the end points $\mb_0$ and  $\mb$, but also on the equivalence class of $\mathcal{C}$.}. Note that $A$ is linear homogeneous in $D$ and $\mub$ which, for the expansion near $D=\mub=0$, corresponds to a certain generalization of the canonical form of Ref. \cite{Henn2013} to the case of analytical regularization. Thanks to this property,  the expansion around the point $D=\mub=0$ has the form
\begin{equation}\label{eq:pexpe}
U(-2\e,-2\boldsymbol{\tau},\mb)=\sum_{n\geqslant0,\boldsymbol{k}\geqslant0} C_{n,\boldsymbol{k}}(\mb) \e^n\prod_{l} \tau_l^{k_l}\,,
\end{equation}
where the coefficients $C_{n,\boldsymbol{l}}(\mb)$ are  the iterated integrals expressed in terms of the Goncharov's polylogarithms \cite{Goncharov1998,Weinzierl:2002hv}. 

Let us note that the recurrence relations obtained in the previous section allow us to express via the same polylogarithmic functions also the expansions near any integer $D$ and even $\mub$. In particular, we can do it for the expansion near any odd integer value of $D$ and integer exponents $\boldsymbol{\alpha}$. 

\begin{figure}
	\centering
	\includegraphics[width=0.5\linewidth]{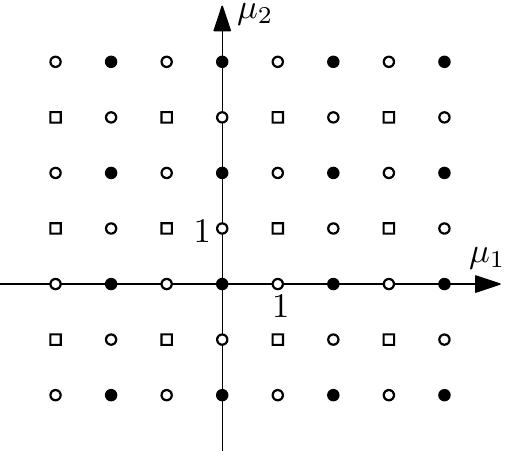}
	\caption{$\mu_1-\mu_2$ plane at fixed integer $D$. The quadratic relations exist near any marked point (disk, circle or square). Solid disks correspond to the points where, in addition, the expansion can be expressed via polylogarithms. Depending on whether $D$ is even or odd, the disks correspond to half-integer or integer powers of the propagators. When $D$ is even, integer powers of propagators correspond to empty squares.}
	\label{fig:muplane}
\end{figure}

In the points where at least one of $\mu_l$ is an odd number, the recurrence relations can not be used to express the corresponding expansion via polylogarithms. However, as we shall see soon, they can be used to obtain the quadratic relations for the expansion coefficients, see Fig. \ref{fig:muplane}. These relations are closely related to the ones described in Ref. \cite{Lee2018} in the case when matrix $A$ is symmetric and  proportional to $\epsilon$ or $\epsilon+1/2$. 

In order to obtain these relations, we first note that the matrix $A(D,\mub)$ has the following symmetry
\begin{equation}\label{eq:A_symmetry}
A^{\intercal}W=WA\,.
\end{equation}
Let $F_{\pm}$ be the two solutions of the equations 
\begin{equation}
dF_{\pm}=\pm A F_{\pm}\,.
\end{equation}
Then, from Eq. \eqref{eq:A_symmetry} it follows that the bilinear form $F^{T}_-WF_+$ is independent of $\mb$:
\begin{equation}
d(F^{\intercal}_-WF_+)=-F^{\intercal}_-A^{\intercal}WF_++F^{\intercal}_-WAF_+ = 0\,.
\end{equation}

Now we note that $A(-D,-\mub)=-A(D,\mub)$ and, therefore, $F_-(D,\mub)$ obeys the same equation as $F_+(-D,-\mub)$. So, given two solutions $T_{1,2}(D,\mub,\mb)$ of Eq. \eqref{eq:dlog}, we can construct the conserved bilinear form
\begin{equation}\label{eq:qrel}
T_{1}^{\intercal}(-D,-\mub,\mb) W(D,\mub)T_{2}(D,\mub,\mb) = C_{12}(D,\mub)\,,
\end{equation}
where $C_{12}(D,\mub)$ in the right-hand side denotes some quantity which, in general, depends on $D$ and $\mub$, but not on $\mb$.

Now let $T_{1,2}$ also satisfy the recurrence relations \eqref{eq:alphashift} and \eqref{eq:Dshift}. Suppose that we are interested in the expansion near $D=D^{\star}$ and $\mub=\mub^{\star}$, where $2D^{\star}$ and $\mu_l^{\star}$ are integer. Then we can write 
\begin{equation}\label{eq:qrel1}
T_{1}^{\intercal}(D^\star+2\epsilon,\mub^\star+2\boldsymbol{\tau},\mb) B T_{2}(D^\star-2\epsilon,\mub^\star-2\boldsymbol{\tau},\mb)= C_{12}(D^\star-2\epsilon,\mub^\star-2\boldsymbol{\tau})\,,
\end{equation}
where $C_{12}(D,\mub)$ is the same function as in Eq. \eqref{eq:qrel} and 
\begin{multline}
B=B(D^\star,\mub^\star|\,\epsilon,\boldsymbol{\tau},\mb)=
\prod_{l=0}^{L}
\left(
\prod_{\nu=1}^{\mu_l^\star} R_l^{\intercal}(D^\star+2\epsilon,-\mub^\star+2\sum _{j=l}^L\mu_j^\star\boldsymbol{e}_j-2\nu\boldsymbol{e}_l+2\boldsymbol{\tau})
\right)\\
\times\left[\prod_{d=1}^{2D^\star} R^{-1\intercal}(D^\star-d+2\epsilon,-\mub^\star+2\boldsymbol{\tau})\right] W(D^\star-2\epsilon,\mub^\star-2\boldsymbol{\tau})\,.
\end{multline}
Using the commutation relations derived in the Appendix \ref{sec:Compatibility} one can show that $B$ has the symmetry
\begin{equation}\label{eq:Bsym}
B^{\intercal}(\epsilon,\boldsymbol{\tau})=(-1)^{2D^\star+\sum_l\mu_l^\star+1} B(-\epsilon,-\boldsymbol{\tau})\,.
\end{equation}
When expanded in $\epsilon$ and $\boldsymbol{\tau}$, the identity \eqref{eq:qrel1} provides a lot of quadratic constraints for the expansion coefficients of the solutions near $D=D^{\star}$ and $\mub=\mub^{\star}$. 
Equations \eqref{eq:qrel} and   \eqref{eq:qrel1} are important results of the present paper.

Let us now specialize Eq. \eqref{eq:qrel} to the case $T_1=V^{(\varrho_0,\boldsymbol{\rho})}$, $T_2=V^{(\tilde\varrho_0,\boldsymbol{\tilde\rho})}$. 
In order to calculate the constant in the right-hand side, we consider the limit $m_l\to 0$ ($l=1,\ldots, L$) at fixed $m_0$. In this limit the main contribution comes from the region $m_0x\sim1$ and from Eq. \eqref{eq:Lauricella} we see that
\begin{equation}\label{eq:Vasym}
	V_{\boldsymbol{a}}^{(\varrho_0,\boldsymbol{\rho})}(D,\mub)=m_0^{\mu_0-D}\prod_{l=1}^L m_0^{\mu_l} \left(\frac{m_l}{m_0}\right)^{\mu_l(1-\rho_l)+1-\delta_{a_l\rho_l}} 
	F\,,
\end{equation}
where $F$ is a function of $({m_l}/{m_0})^2$, analytic and nonzero at the origin. The value of $F$ at the origin can be obtained by replacing $F_C^{(L)}\to 1$ in Eq. \eqref{eq:Lauricella}.
Since we know that the right-hand side of Eq. \eqref{eq:qrel} does not depend on masses, it can be nonzero only if $\boldsymbol{\tilde\rho}=\boldsymbol{\rho}$. Indeed, the power $m_l^{\mu_l(\tilde{\rho}_l-{\rho}_l)}$ can not be canceled if $\tilde{\rho}_l\neq \rho_l$. Besides, the exponent $1-\delta_{a_l\rho_l}$ in Eq. \eqref{eq:Vasym} shows that only the contribution of the components with $a_{l}=\rho_l$ ($l=1,2,\ldots,L$) may contribute to the limit $m_l\to 0$. Taking the integral over $x$, we have
\begin{multline}\label{eq:Vasym1}
V_{a_0\boldsymbol{\rho}}^{(\varrho_0,\boldsymbol{\rho})} \simeq \delta_{a_0\varrho_0}\left(-1\right)^{a_{0}}2^{D-1}m_{0}^{\mu_{0}-D}\prod_{l=1}^{L}(-1)^{\rho_l}\Gamma\left[\tfrac{1}{2}\left(1-\mu_l(-1)^{\rho_l} \right)
\right] m_0^{\mu_l\rho_l}m_{l}^{\mu_{l}\left(1-\rho_{l}\right)}\\
\times \Gamma\left[\tfrac{1}{2}\left(D-\mu_0a_{0}-\textstyle\sum_{l=1}^{L}\mu_{l}\rho_{l}\right)\right]	\Gamma\left[\tfrac{1}{2}\left(D+1-\mu_{0}\bar{a}_{0}-\textstyle\sum_{l=1}^{L}\mu_{l}\rho_{l}\right)\right]
\end{multline}

Using this asymptotics in the left-hand side of Eq. \eqref{eq:qrel}, we fix the right-hand side of this identity. We find
\begin{multline}\label{eq:quadconstraint}
V^{(\tilde\varrho_0\boldsymbol{\tilde\rho})\intercal}\left(-D,-\boldsymbol{\mu}\right)W\left(D,\boldsymbol{\mu}\right)V^{(\varrho_0,\boldsymbol{\rho})}\left(D,\boldsymbol{\mu}\right)
=
\frac12\delta_{{\tilde\varrho_0}{\varrho_0}}
\delta_{\boldsymbol{\tilde\rho}\boldsymbol{\rho}}\left[\prod_{l=1}^{L}\frac{\pi}{\cos\tfrac{\pi\mu_{l}}{2}}\right]\\ \times
\frac{\pi}{\sin\tfrac{\pi}{2}\left(D-\mu_{0}\varrho_{0}-\sum_{l=1}^{L}\mu_{l}\rho_{l}\right)}\frac{\pi}{\cos\tfrac{\pi}{2}\left(D-\mu_{0}\bar\varrho_{0}-\sum_{l=1}^L\mu_{l}\rho_{l}\right)}
\end{multline}
Recall that $\bar\varrho_{0}=1-\varrho_{0}$. Let us note that in Ref. \cite{Goto2013} some bilinear relations between Lauricella functions $F_C$ have been derived using different methods. It would be interesting to compare them with Eq. \eqref{eq:quadconstraint}.

Summing over $\varrho_0$ and  $\tilde\varrho_0$ and using the identity
\begin{equation}\label{eq:id1}
	\frac{1}{\sin\tfrac{x}{2}}\frac{1}{\cos\tfrac{x-y}{2}}+\frac{1}{\sin\tfrac{x-y}{2}}\frac{1}{\cos\tfrac{x}{2}}
	=\frac{2}{\cos\tfrac{y}{2}}\left[\frac{1}{\sin x}+\frac{1}{\sin(x-y)}\right]\,,
\end{equation}
we obtain
\begin{equation}\label{eq:quadconstraint1}
	V^{(\boldsymbol{\tilde\rho})\intercal}\left(-D,-\boldsymbol{\mu}\right)W\left(D,\boldsymbol{\mu}\right)V^{(\boldsymbol{\rho})}\left(D,\boldsymbol{\mu}\right)=
	\delta_{\boldsymbol{\tilde\rho}\boldsymbol{\rho}}\left[\prod_{l=0}^{L}\frac{\pi}{\cos\tfrac{\pi\mu_{l}}{2}}\right]\sum_{\rho_{0}=0,1}\frac{\pi}{\sin\pi\left(D-\sum_{l=0}^{L}\mu_{l}\rho_{l}\right)}\,.
\end{equation}
Finally, summing over $\boldsymbol{\rho}$ and  $\boldsymbol{\tilde\rho}$ and using Eq. \eqref{eq:TviaV} we have
\begin{equation}
T^{\intercal}\left(-D,-\boldsymbol{\mu}\right)W\left(D,\boldsymbol{\mu}\right)T\left(D,\boldsymbol{\mu}\right)=
\left[\prod_{l=0}^{L}\frac{\pi}{\cos\tfrac{\pi\mu_{l}}{2}}\right]\sum_{\rho_{0}=0,1}\ldots \sum_{\rho_{L}=0,1}\frac{\pi}{\sin\pi\left(D-\sum_{l=0}^{L}\mu_{l}\rho_{l}\right)}\,.
\end{equation}
Note that we can obtain the bilinear relation separately for $\mathcal{P}$-even part $T_+=\frac{1+\mathcal{P}}{2}T$. In order to do this, in addition to Eq. \eqref{eq:id1},  we also use the identity
\begin{equation}
\frac{1}{\sin\tfrac{x}{2}}\frac{1}{\cos\tfrac{x-y}{2}}-\frac{1}{\sin\tfrac{x-y}{2}}\frac{1}{\cos\tfrac{x}{2}}
=\frac{2}{\cos\tfrac{y}{2}}\left[\cot x-\cot(x-y)\right]\,.
\end{equation}
We obtain
\begin{multline}
T^{\intercal}_+\left(-D,-\boldsymbol{\mu}\right)W\left(D,\boldsymbol{\mu}\right)T_+\left(D,\boldsymbol{\mu}\right)=
\left[\prod_{l=0}^{L}\frac{\pi}{\cos\tfrac{\pi\mu_{l}}{2}}\right]\sum_{\rho_{0}=0,1}\ldots \sum_{\rho_{L}=0,1}\\
\times\frac\pi 2(-1)^{\sum_{l=0}^{L}\rho_l}\cot\frac{\pi}{2}\left[D-\sum_{l=0}^L(\mu_l-1)\rho_l\right]\,.
\end{multline}

\section{Special cases when the system acquires a block-triangular form}

\subsection{Removing analytic regularization}

Let us first consider the special case $\mu_l=\mu=D-1$, which corresponds to the removal of the analytical regularization. The differential system \eqref{eq:dlog} in this case should have a block-triangular form, corresponding to the decoupling of the trivial ``clover-leaf'' integrals arising from the contraction of one of $L+1$ lines. In order to see this structure, it is convenient to pass to the new functions 
\begin{equation}\label{eq:Yfunctions}
	Y(\mu,\mb)=M\, T(D,\mub,\mb)\bigg\vert_{\substack{D=\mu+1 \\ \mu_l=\mu}}
\end{equation}
The differential system for these functions has the form
\begin{align}\label{eq:Yequation}
	dY&=\mu HY\,,\\
	\label{eq:H}
	H&=\left(N-1\right)\frac{dM}{M}+\sum_{l=0}^L\bar n_{l}\frac{dm_{l}}{m_{l}}\,,
\end{align}
where $N=\sum_{l=0}^L n_{l}$. Note that Eq. \eqref{eq:Yequation} corresponds to $(\e+1/2)$-form discussed in Ref. \cite{Lee2018}. In particular, this form allows one to write down the $\epsilon$-expansion of the solution near $D=1$ in terms of Goncharov polylogarithms (see Appendix \ref{sec:D=1} for details).
Let us observe that the equations for each of the components 
\begin{equation}\label{eq:clover_integrals}
	Y_{10\ldots0},\,Y_{01\ldots0},\ldots,Y_{0\ldots01}
\end{equation}
decouple and acquire the form
\begin{equation}
	d Y_{\ldots 0\underset{l}{1}0\ldots}= \mu\left[\sum_{i\neq l}\frac{dm_{i}}{m_{i}}\right] Y_{\ldots 0\underset{l}{1}0\ldots}
\end{equation}
with the general solution being
\begin{equation}
Y_{\ldots 0\underset{l}{1}0\ldots}\propto \prod_{i\neq l}m_{i}^\mu\,.
\end{equation}
These are just the equations for the clover-leaf integrals. One can check explicitly that $Y_{10\ldots0},\,Y_{01\ldots0},\ldots,Y_{0\ldots01}$ are indeed evaluated to the corresponding expressions for the clover-leaf integrals after substitution of Eq. \eqref{eq:T_a} into Eq. \eqref{eq:Yfunctions}. The simplest way to check this is to represent $Y(\mu)$ as
\begin{equation}\label{eq:Yaslim}
	Y(\mu)=\lim_{D\to\mu+1} W(D-1,\mu) R^{-1}(D-1,\mu) T(D,\mu)=\lim_{D\to\mu+1} W(D-1,\mu) T(D-1,\mu)
\end{equation}
and to note that the corresponding elements of $W$ are vanishing in this limit. Then, only the pole part of the corresponding components of $T(D-1,\mu)$ have to be evaluated (note that $W$ is a diagonal matrix).

Let us consider now the homogeneous differential system obtained by putting components \eqref{eq:clover_integrals} to zero. This is the system for the maximally cut tadpole diagram. It has the same form \eqref{eq:Yequation}, where now the action of $H$  is restricted to the subspace with
\begin{equation}\label{eq:constraints}
	Y_{10\ldots0}=Y_{01\ldots0}=\ldots =Y_{0\ldots01}=0\,.
\end{equation}

 Note that in this subspace the operator $N-1$ is invertible. From now on to the end of this subsection, we assume that all operators are restricted to this subspace and that all vector functions belong to it. Then, using the property
\begin{equation}
	H^{\intercal} (N-1)^{-1} = (N-1)^{-1}H\,,
\end{equation}
and following the same path as in the previous section, we obtain the bilinear constraint
\begin{equation}
	Y_{1}^{\intercal}(-\mu,\mb)  (N-1)^{-1} Y_{2}(\mu,\mb) = \mathrm{const}
\end{equation}
Let us now determine the basis of solutions in the subspace constrained by Eq. \eqref{eq:constraints}. We define
\begin{equation}
Y^{(\varrho_0,\boldsymbol{\rho})}(\mu,\mb)=M\, V^{(\varrho_0,\boldsymbol{\rho})}(D,\mub,\mb)\bigg\vert_{\substack{D=\mu+1 \\ \mu_l=\mu}}\,.
\end{equation}
One can check explicitly that $2^{L+1}-L-1$ such functions with $\varrho_0\neq|\boldsymbol{\rho}|$ satisfy the constraints \eqref{eq:constraints} and, therefore, form the basis  we are looking for\footnote{In particular, when $|\boldsymbol{\rho}|\geqslant 2$, the function $Y^{(\boldsymbol{\rho})}=M\, V^{(\boldsymbol{\rho})}\big\vert_{\substack{D=\mu+1 \\ \mu_l=\mu}}=Y^{(0,\boldsymbol{\rho})}+Y^{(1,\boldsymbol{\rho})}$ is the homogeneous solution.\label{ft:Yrho}}.

For two basis functions, $Y^{(\tilde\varrho_0,\boldsymbol{\tilde\rho})\intercal}$ and $Y^{(\varrho_0,\boldsymbol{\rho})}$, using Eqs. \eqref{eq:quadconstraint} and \eqref{eq:Yaslim}, we have
\begin{multline}\label{eq:Yqrel}
Y^{(\tilde\varrho_0,\boldsymbol{\tilde\rho})\intercal}\left(-\mu\right)\left[\mu(N-1)\right]^{-1}Y^{(\varrho_0,\boldsymbol{\rho})}\left(\mu\right)
=
\frac{1}2\delta_{{\tilde\varrho_0}{\varrho_0}}
\delta_{\boldsymbol{\tilde\rho}\boldsymbol{\rho}}\left(\frac{\pi}{\cos\tfrac{\pi\mu}{2}}\right)^L\\ \times
\frac{\pi}{\sin\tfrac{\pi \mu}{2}\left(\varrho_{0}-|\boldsymbol{\rho}|\right)}\frac{\pi}{\cos\tfrac{\pi\mu}{2}\left(\bar\varrho_{0}-|\boldsymbol{\rho}|\right)}\,.
\end{multline}
Note that the right-hand side of this identity only makes sense when $\varrho_0\neq|\boldsymbol{\rho}|$, otherwise the argument of $\sin$ function in the denominator becomes zero. 

In order to obtain the quadratic relations for the coefficients of $\e$-expansion near $\mu=\mu^\star$ for some integer $\mu^\star$, we have to substitute $\mu=\mu^\star-2\e$ in Eq. \eqref{eq:Yqrel} and use the relation
\begin{equation}\label{eq:Yshift}
Y^{(\tilde\varrho_0,\boldsymbol{\tilde\rho})\intercal}\left(-\mu^{\star}+2\e\right)
=Y^{(\tilde\varrho_0,\boldsymbol{\tilde\rho})\intercal}\left(\mu^{\star}+2\e\right)\left[(-1)^{(L+1)\mu^\star}\prod_{\nu=1}^{\mu^\star} \mathfrak{R}(-2\nu+\mu^\star+2\e)\right]\,,
\end{equation}
where $\mathfrak{R}(\mu)=\mathfrak{R}(D,\mub)\big\vert_{\substack{D=\mu \\ \mu_l=\mu}}$ is the operator of Eq. \eqref{eq:Dmushift}  restricted to the subspace \eqref{eq:constraints}.

Note that the quadratic relations obtained by substituting Eq. \eqref{eq:Yshift} into Eq. \eqref{eq:Yqrel} are valid only for the solutions of homogeneous equations. A natural question arises: whether one can obtain similar relations also for generic solutions of inhomogeneous equations? Within our approach, the answer is negative. The bilinear relations \eqref{eq:qrel} are valid for any two solutions $T_1$ and $T_2$ even if we put $D-1=\mu_l=\mu$. However, in order to obtain the quadratic relations near any integer point $\mu=\mu^\star\geqslant0$, we have to shift the first argument in the left factor $T_1^{\intercal}$ at least twice by the operator $R^{-1}$. Then the second operator $R^{-1}(-\mu,-\mu)$ does not make sense (since $R(-\mu,-\mu)=-\mu M^{-1}(N-1)$ is not invertible).

\subsection{Groups of identical lines}
Let us now discuss the case when some lines share the same $\mu$ and $m$. Then we can write
\begin{equation}\label{eq:tadpole-xspace1}
\mathcal{T}
=\intop_0^{\infty} d x\,x^{D-1} \prod_{k=0}^{K}\left[P_0(\mu_k,m_k,x)\right]^{r_k}\,,
\end{equation}
where $r_k$ is the number of lines with mass $m_k$ and parameter $\mu_k$, so that $\sum_k r_k=L+1$. There is a symmetry group $\cal G$ generated by the permutations of lines sharing the same mass and parameter. This symmetry, in particular, holds for the matrix $A$ in the right-hand side of the differential system, which now can be written as 
\begin{gather}
A = dM\,M^{-1}W + \sum_{k=0}^{K} \frac{\mu_{k}}{2}\left(r_k+2S_{zk}\right)\frac{dm_k}{m_k} \\
M=2\sum_{k=0}^{K} m_k S_{xk}, \quad W=\sum_{k=0}^{K} \frac{\mu_{k}}{2}\left(r_k-2S_{zk}\right)-D\,,
\end{gather}
where
\begin{equation}
	S_{xk}=\frac12\sum_{l=u_k}^{u_k+r_k-1} \sigma_{xl}\,,\quad
	S_{zk}=\frac12\sum_{l=u_k}^{u_k+r_k-1}\sigma_{zl}\,.
\end{equation}
Here $u_k=\sum_{s=0}^{k-1} r_s$, and the sequence $u_{k},u_{k}+1,\ldots, u_{k}+r_k-1$ enumerates the lines in $k$-th group. It is remarkable that $A$ is expressed solely via $S_{xk}$ and $S_{zk}$ which are among the standard generators of the Lie algebra ${\mathfrak{sl}}(2,\mathbb{C})$. We can express via the same generators other operators which we have used previously, in particular $\mathcal{P} = i^{L+1}\prod_{k=0}^{K} e^{-i\pi S_{zk}}$.

Note that there is a subtle point here that we want to stress. When all $r_k$ are even, the matrix $M$ obviously has zero eigenvalue and, therefore, is not invertible. Thus the matrix $A$ is ill-defined. So, from now on we assume that at least one $r_k$ is odd, i.e., at least one group contains odd number of lines. 

The symmetry of the matrix $A$ leads to the splitting of the system into separate subsystems, each corresponding to a specific irreducible representation of $\cal G$. Since the tadpole integral \eqref{eq:tadpole-xspace1} is invariant under the action of this group, we are mostly interested in the subspace spanned by the tensors $T_{\boldsymbol{a}}$ symmetric with respect to all permutations of indices from this symmetry group.

Alternatively, we can reduce the number of auxiliary functions from $2^{L+1}=2^{\sum_kr_k}$ to $\prod_k(r_k+1)$ from the very beginning. Namely, we can introduce the functions
\begin{equation}\label{eq:T_a1}
	T_{\boldsymbol{a}}=\intop_0^{\infty} d x\,x^{D-1} \prod_{k}{r_k \choose a_k}^{1/2}\left[P_0(\mu_k,m_k,x)\right]^{r_k-a_k}\left[P_1(\mu_k,m_k,x)\right]^{a_k}
\end{equation}
where $\boldsymbol{a}=a_0\ldots a_K$ and $a_k$ runs from $0$ to $r_k$, and ${n \choose k}= \frac{n!}{k!(n-k)!}$ is the binomial coefficient.

As before, it is convenient to treat the quantities $T_{\boldsymbol{a}}$ as components of the vector $T$ in $\mathbb{C}^{r_0+1}\otimes\ldots \otimes \mathbb{C}^{r_K+1}$. Then the operators $S_{xk}$ and $S_{zk}$ act on $k$-th factor as usual spin operators, i.e., the matrices with nonzero elements being
\begin{align}
	\left(S_{xk}\right)_{l,l-1}&=\left(S_{xk}\right)_{l-1,l}=\frac12\sqrt{l(r_k+1-l)}\quad \text{($l=1,\ldots r_k$)}\,,\\
	\left(S_{zk}\right)_{ll}&=r_k/2-l\quad \text{($l=0,\ldots r_k$)}\,.
\end{align}
Again, thanks to the parity operator the differential system consisting of $\prod_k(r_k+1)$ equations splits into two subsystems corresponding to $\mathcal{P}=+1$ and  $\mathcal{P}=-1$, the first involving components with $|\boldsymbol{a}|=\sum_{k}a_k$ even, the second with $|\boldsymbol{a}|$ odd. Note that since we assume that at least one $r_k$ is odd, the numbers of even and odd components coincide and are equal to $\frac12\prod_k(r_k+1)$.

Let us now discuss the parameter and dimension shifting relations. Here we have to take into account that shifting separately each parameter is not compatible with the permutation symmetry. Fortunately, for the dimension shift at fixed $\alpha_k$ it is sufficient to have possibility to shift all parameters simultaneously. Therefore, we need to define the action of 
\begin{equation}\label{eq:calRk}
	\mathcal{R}_k=\prod_{l=u_k}^{u_k+r_k-1} R_l\left(D,\mub-2\textstyle\sum_{j=l+1}^{u_k+r_k-1}\boldsymbol{e}_j\right)
\end{equation} 
on the symmetric subspace. After some transformations we obtain
\begin{equation}
	\mathcal{R}_k=\left(\frac{i}{2m_{k}}\right)^{r_k}e^{i\pi S_{xk}}\sum_{n=0}^{r_k}\frac1{n!}\left(\frac{\mu_k-1}{m_k}S_{-k}\right)^n \prod_{l=-r_k+n+1}^{0} R(D+l)\,,
\end{equation}
where $S_{-k}=S_{xk}-iS_{yk}=S_{xk}+[S_{xk},S_{zk}]$ which is the matrix with nonzero elements being
\begin{equation}
\left(S_{-k}\right)_{l,l-1}=\sqrt{(r_k+1-l)l}\quad \text{($l=1,\ldots r_k$)}\,.
\end{equation}

Let us now construct the basis of solutions. We remind that we restrict ourselves to the case when at least one of $r_k$ is odd. We assume that $r_0$ is odd. Similarly to Section \ref{sec:basis}, we want to replace in Eq. \eqref{eq:tadpole-xspace1} some $P_0$ with $Q_0^{(0)}$ or  $Q_0^{(1)}$ defined in Eq. \eqref{eq:Qdef}. Assuming that $m_0>\sum_{k=1}^{K} r_k m_k$, for convergence of the integral over $x$ we keep $[P_0(\mu_0,m_0,x)]^{\lceil r_0/2\rceil}$ and replace all other $P_0$ with $Q_0^{(0)}$ or  $Q_0^{(1)}$. Thus we have
\begin{multline}
	V_{\boldsymbol{0}}^{(\rho_0\boldsymbol{\rho})}=
	\intop_0^{\infty} d x\,x^{D-1}
	\left[P_0(\mu_0,m_0,x)\right]^{\lceil r_0/2\rceil}\left[Q_0^{(0)}(\mu_0,m_0,x)\right]^{\lfloor r_0/2\rfloor-\rho_0}\left[Q_0^{(1)}(\mu_0,m_0,x)\right]^{\rho_0}\\
	\times \prod_{k=1}^{K}\left[Q_0^{(0)}(\mu_k,m_k,x)\right]^{r_k-\rho_k}
	\left[Q_0^{(1)}(\mu_k,m_k,x)\right]^{\rho_k}\,,
\end{multline}
where  $\rho_0=0,1,\ldots, \lfloor r_0/2\rfloor $, $\rho_{k}=0,1,\ldots, r_k$ ($k>0$), and $\boldsymbol{\rho}=\rho_1,\ldots \rho_K$. Note that since $r_0$ is odd, we have $\lceil r_0/2\rceil=(r_0+1)/2$,  $\lfloor r_0/2\rfloor=(r_0-1)/2$. Then there are \linebreak $(\lfloor r_0/2\rfloor+1)\prod_{k=1}^K(r_k+1)=\tfrac12\prod_{k=0}^K(r_k+1)$ functions $V_{\boldsymbol{0}}^{(\boldsymbol{\rho})}$. This is exactly the correct dimension of $\mathcal{P}$-even part of solution space. If we modify similarly other components of $T_{\boldsymbol{a}}$ from Eq. \eqref{eq:T_a1}, we immediately see that we have exactly $\tfrac12\prod_{k=0}^K(r_k+1)$ basis vectors also in $\mathcal{P}$-odd part of solution space, which also coincides with the correct dimension of this subspace. Note that the components other than $T_{\boldsymbol{0}}$ involve proper symmetrization within each group of equivalent lines. For the sake of completeness we present here the result:
\begin{multline}\label{eq:Vrho1}
V_{\boldsymbol{a}}^{(\rho_0\boldsymbol{\rho})}=
\intop_0^{\infty} d x\,x^{D-1}\sum_{b_0,c_0} 
\frac{{\lceil r_0/2\rceil \choose c_0}{\rho_0\choose b_0}
{\lfloor r_0/2\rfloor-\rho_0\choose a_0-b_0-c_0}}{{r_0\choose a_0}^{1/2}}\\
\times
 \left(P_0^{\lceil r_0/2\rceil-c_0}P_1^{c_0}\left[Q_0^{(0)}\right]^{\lfloor r_0/2\rfloor-\rho_0-a_0+b_0+c_0}\left[Q_1^{(0)}\right]^{a_0-b_0-c_0}\left[Q_0^{(1)}\right]^{\rho_0-b_0}\left[Q_1^{(1)}\right]^{b_0}\right)(\mu_0,m_0,x)\\
\times \prod_{k=1}^{K}
\sum_{b_k} 
\frac{{\rho_k\choose b_k}
	{r_k-\rho_k\choose a_k-b_k}}{{r_k\choose a_k}^{1/2}}
\left(\left[Q_0^{(0)}\right]^{r_k-\rho_k-a_k+b_k}\left[Q_1^{(0)}\right]^{a_k-b_k}\left[Q_0^{(1)}\right]^{\rho_k-b_k}\left[Q_1^{(1)}\right]^{b_k}\right)(\mu_k,m_k,x)\,.
\end{multline}

Again, we have the quadratic relations of the form
\begin{equation}\label{eq:qrelS}
V^{(\tilde{\rho}_0\boldsymbol{\tilde\rho})\intercal}\left(-D,-\boldsymbol{\mu}\right)W\left(D,\boldsymbol{\mu}\right)V^{(\rho_0\boldsymbol{\rho})}\left(D,\boldsymbol{\mu}\right)\\
=\delta_{\boldsymbol{\tilde\rho}\boldsymbol{\rho}}C(\rho_0,\tilde{\rho}_0,\boldsymbol{\rho}|D,\mub)\,.
\end{equation}
If $r_0=1$, the function $C(\rho_0,\tilde{\rho}_0,\boldsymbol{\rho}|D,\mub)$ in the right-hand side can be easily deduced from Eq. \eqref{eq:quadconstraint1}: 
\begin{equation}\label{eq:qrelrhs}
	C(0,0,\boldsymbol{\rho}|D,\mub)=\left[\prod_{k=0}^{K}\left(\frac{\pi}{\cos\tfrac{\pi\mu_{k}}{2}}\right)^{r_k}\right]\sum_{\rho_0=0,1}\frac{\pi}{\sin\pi\left(D-\sum_{k=0}^{K}\mu_{k}\rho_{k}\right)}\,,\quad (r_0=1)
\end{equation}
However, it is not quite clear how to fix this function for $r_0>1$, and we leave this question for further investigations. The reason why the case $r_0=1$ is simple for us is because, while constructing the basis of solutions,  we always assumed that one mass ($m_0$) is larger than the sum of all others. Therefore we did not have obligations to consider the questions of analytical continuation of the obtained solutions across (or around) the singular hypersurfaces defined by Eq. \eqref{eq:singular}. However, if $r_0>1$, we would have to do it.

Finally, we note that if we remove the analytical regularization, the $\mathcal{P}$-even subsystem contains $K+1$ decoupled equations and the number of its homogeneous solutions becomes $\frac12\prod_k(r_k+1)-K-1$.

\subsection{Removing dimensional regularization at $D=2$}

\begin{figure}
	\centering
	\includegraphics[width=0.4\linewidth]{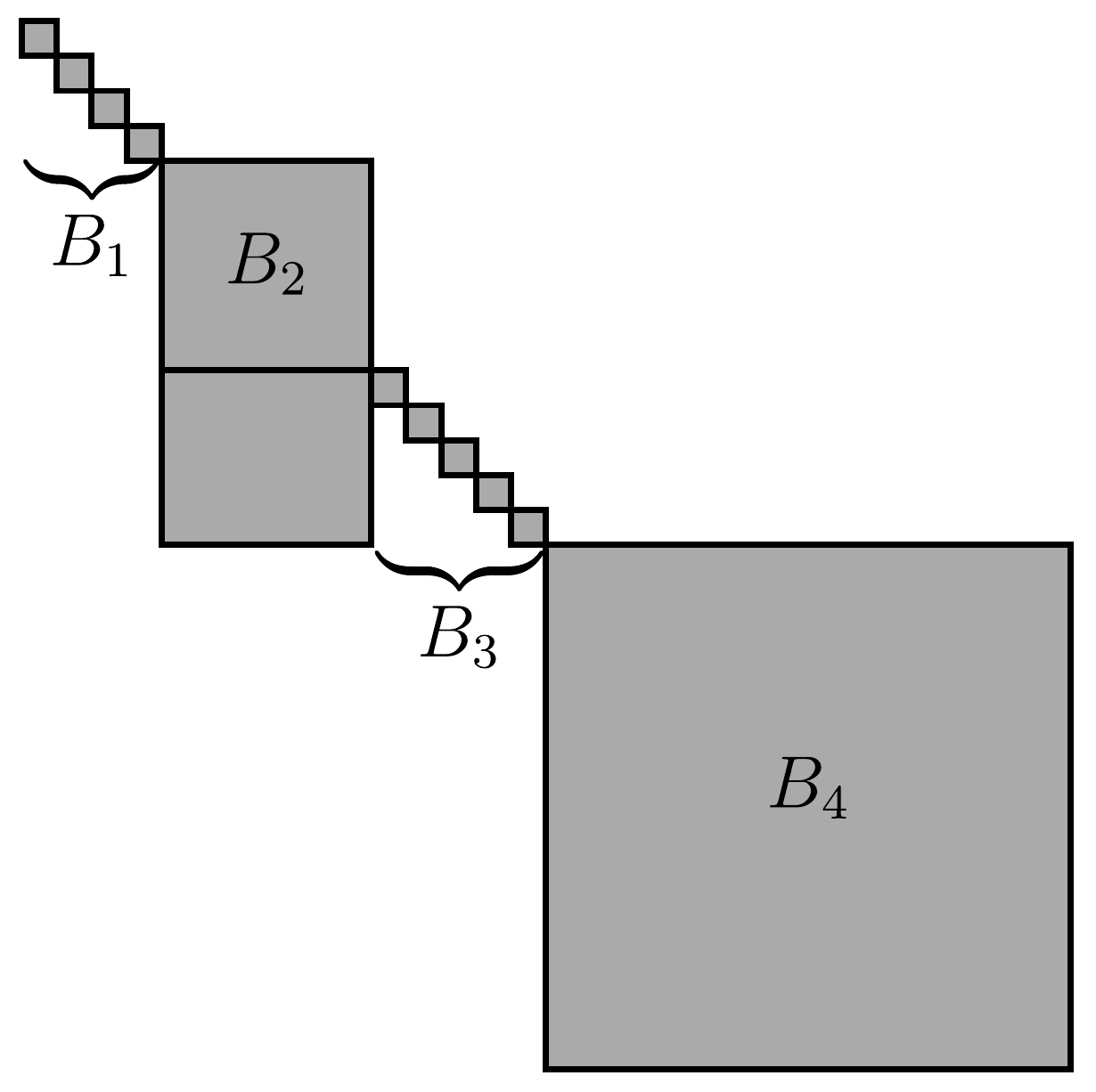}
	\caption{ Block-triangular structure of the matrix in the right-hand side of the differential system for $\tilde T$. The block $B_1$ is a diagonal matrix corresponding to $L+1$ entries of the form $Y_{\ldots 010\ldots}$. The block $B_2$ corresponds to $2^L-(L+1)L/2$ entries $T_{\boldsymbol{a}}\ (|\boldsymbol{a}| \text{ even and }|\boldsymbol{a}| \neq 2)$. The block $B_3$ is a diagonal matrix corresponding to $(L+1)(L-2)/2$ remaining entries of $T_{\boldsymbol{a}}$ with $|\boldsymbol{a}| = 2$. The block $B_4$ corresponds to $2^L$ entries $T_{\boldsymbol{a}}$ with odd $|\boldsymbol{a}|$.
	}
	\label{fig:splitting}
\end{figure}

Let us now discuss some important features of the obtained results at $D=\mu+1=2$. Below we assume that $L>1$. The matrix $A$ in this case has the form
\begin{equation}
	A=dM\,M^{-1}(N-2) + \sum_{l=0}^{L}\bar{n}_{l}\frac{dm_l}{m_l}  
\end{equation}
The second term is diagonal and it is clear that the equations for the components with $|\boldsymbol{a}|\neq 2$ form a closed subsystem. On the other hand, if we pass to $Y$, Eq. \eqref{eq:Yfunctions}, we obtain the system
\begin{equation}
dY=HY
\end{equation}
with $H$ defined in Eq. \eqref{eq:H}. In this form, the equations for $Y_{\boldsymbol{a}}$ components with $|\boldsymbol{a}|=1$ decouple. Let us consider the linear transformation of column of functions $T_{\boldsymbol{a}}$ which replaces $L+1$ entries $T_{110\ldots 0},\,T_{1010\ldots 0},\ldots ,\,T_{10\ldots 10},\,T_{10\ldots 01}$, and $T_{0110\ldots 0}$ with $Y_{100\ldots 0},\,Y_{010\ldots 0},\ldots ,Y_{0\ldots 01}$. It is easy to check that this transformation is non-degenerate. With the new column of functions $\tilde T$ the differential system acquires the block-triangular form depicted in Fig. \ref{fig:splitting}.
Let us observe that the component $T_{\boldsymbol{0}}$ enters the closed system of $2^L-(L+1)L/2$ equations determined by the block $B_2$. Therefore, in a certain sense, we find that the number of master integrals at $D=2$ is equal to $2^L-(L+1)L/2$. This is in agreement with known cases $L\leqslant 4$, \cite{Adams2013,TancrediPC2019}. 

Let us comment about the number of master integrals at $D=2$ when there are groups of identical masses. In the notations of previous subsection, we obtain that the number of master integrals is equal to $\frac12\prod_k(r_k+1)-(K+1)(K+2)/2+\sum_k\delta_{r_k,1}$. In particular, for $L$-loop sunrise graph with equal masses we have $L+2-1-1 =L$ master integrals.
\paragraph{Basis of solutions.}
Constructing the basis of solution space at $D=2$ appears to be extremely tricky and deserves a dedicated consideration elsewhere.  Here we will only explain the complications that arise on the way.

First, one might be tempted to pass to the basis of functions $V^{[\boldsymbol{\upsilon}]}$ defined as
\begin{equation}
V^{[\boldsymbol{\upsilon}]}=\sum_{\boldsymbol{\rho}}\prod_{l=1}^{L}\left(-\frac{2}{\pi}\sin \pi\e\right)^{{\upsilon}_l}(1-\bar{\rho}_l{\upsilon}_l)V^{(\boldsymbol{\rho})}\,,
\end{equation}
where $\boldsymbol{\upsilon}=\upsilon_1\upsilon_2\ldots \upsilon_L$ is a binary number.
The functions of this basis have the following limit of $\boldsymbol{a}=\boldsymbol{0}$ component:
\begin{equation}\label{eq:V[]}
V^{[\boldsymbol{\upsilon}]}_{\boldsymbol{0}}\stackrel{\epsilon\to 0}{\longrightarrow}\prod_{l=0}^{L}(2m_l) \intop_0^{\infty} dx\,x
K_{0}\left(m_0x\right)
\prod_{l=1}^L \begin{cases}
I_0\left(m_lx\right)\quad \text{if } \upsilon_l=1, \\
K_0\left(m_lx\right)\quad \text{if } \upsilon_l=0
\end{cases}\,.
\end{equation}
However, other components of $V^{[\boldsymbol{\upsilon}]}$ diverge in the limit $\epsilon\to 0$ unless $\boldsymbol{\upsilon} = 1\ldots 1$ because fo the singular asymptotics of $K_1$ function at small argument.

In order to get rid of the divergent overall factor in Eq. \eqref{eq:Vrho} at $\mu\to 1$, let us define the functions 
\begin{equation}
	U^{(\boldsymbol{\rho})}=(-1)^{|\boldsymbol{\rho}|}\left(\frac{2\sin\pi\epsilon}\pi\right)^{L+1}\left. V^{(\boldsymbol{\rho})}\right\vert_{\substack{D=2-2\epsilon \\ \mu_l=1-2\epsilon}}\,.
\end{equation} 
Then if we naively take limit $\epsilon\to 0$ under the integral sign, we can come to a wrong conclusion that all $U^{(\boldsymbol{\rho})}$ tend to one and the same expression independent of $\boldsymbol{\rho}$:
\begin{gather}\label{eq:U_naive_limit}
U^{(\boldsymbol{\rho})}\stackrel{\epsilon\to 0}{\longrightarrow}\prod_{l=0}^{L}(2m_l) \intop_0^{\infty} dx\,x
\begin{pmatrix}
K_{0}\left(m_0x\right)\\
-K_{1}\left(m_0x\right)
\end{pmatrix}
\otimes\bigotimes_{l=1}^L 
\begin{pmatrix}I_0\left(m_lx\right)\\I_1\left(m_lx\right)\end{pmatrix}\,.
\end{gather}
However, Eq. \eqref{eq:U_naive_limit} is correct only for $U^{(\boldsymbol{0})}$. The reason why it breaks down for $\boldsymbol{\rho}\neq \boldsymbol{0}$ is that the expansion of $I_{-1+\epsilon}(mx)$ in $x$ starts from the singular term $\Gamma[\epsilon]^{-1}(mx/2)^{\epsilon-1}$. Although this term is suppressed in $\epsilon$, it may survive after the integration over $x$ due to its singular nature. More precisely, consider the identities
\begin{align}\label{eq:Kint}
	\int dx\,K_{-\epsilon}(x)x^{\alpha-1}	&=2^{\alpha}\Gamma\left[\tfrac{\alpha+\epsilon}{2},\tfrac{\alpha-\epsilon}{2}\right]\,,\nonumber\\
	\int dx\,K_{1-\epsilon}(x)x^{\alpha}	&=2^{\alpha}\Gamma\left[\tfrac{\alpha+\epsilon}{2},\tfrac{\alpha-\epsilon}{2}+1\right]\,.
\end{align}
They show that when $\alpha$ is close to $-n$ ($n=0,1,\ldots$), the integral over $x$ gives poles in $\epsilon$. 
A thorough investigation allows us to establish that nonzero corrections to the naive limit \eqref{eq:U_naive_limit} appear only in the components with $|\boldsymbol{a}|=2$. Moreover, we can explicitly calculate these corrections using Eqs. \eqref{eq:Kint}:
\begin{align}
	{\left[U^{\left(\boldsymbol{\rho}\right)}-U^{\left(\boldsymbol{0}\right)}\right]_{\boldsymbol{a}}}	 &\stackrel{\epsilon\to 0}{\longrightarrow}\delta_{|\boldsymbol{a}|,2}\prod_{l=0}^{L}(2m_l)\left[
	\sum_{\substack{b,c, \\ 1\leqslant b<c\leqslant L}}\frac{a_ba_c\rho_{b}\rho_{c}}{m_bm_c\left|\boldsymbol{\rho}\right|\left(\left|\boldsymbol{\rho}\right|-1\right)}
	-\sum_{b=1}^{L}\frac{a_0a_b\rho_{b}}{m_0m_b\left|\boldsymbol{\rho}\right|}\right]
\end{align}
Here we imply that $\boldsymbol{\rho}\neq \boldsymbol{0}$ and that the first term in square brackets should be omitted when $|\boldsymbol{\rho}|=1$. We can construct the solutions which have nonzero $\epsilon\to 0$ limit exactly in one component with $|\boldsymbol{a}|=2$. Let us adopt the notation $\boldsymbol{1}_{l_1,l_2,\ldots}$ for the vector $\boldsymbol{\rho}$ with $\rho_{l_1}=\rho_{l_2}=\ldots=1$ and other components equal to zero. Then we define
\begin{align}
	U(0,b)&=\lim_{\epsilon\to 0}\left[U^{\left(\boldsymbol{0}\right)}-U^{\left(\boldsymbol{1}_b\right)}\right]\,,&(1\leqslant b\leqslant L)\\
	U(b,c)&=\lim_{\epsilon\to 0}\left[2U^{\left(\boldsymbol{1}_{bc}\right)}-U^{\left(\boldsymbol{1}_b\right)}-U^{\left(\boldsymbol{1}_b\right)}\right]\,,&(1\leqslant b<c\leqslant L)
\end{align}
with the property $[U(b,c)]_{\boldsymbol{a}}=\delta_{\boldsymbol{a},\boldsymbol{1}_{b,c}}\frac{\prod_{l=0}^{L}(2m_l)}{m_b m_c}$ (here $0\leqslant b<c\leqslant L$). These ${L+1 \choose 2}=L(L+1)/2$ solutions are obviously  the solutions of the homogeneous equations for the components with $N=2$. Note that any difference  $U^{\left(\boldsymbol{\rho}\right)}-U^{\left(\boldsymbol{0}\right)}$ tends to a linear combination of $U(b,c)$. Therefore, we construct the combinations that vanish in the limit $\epsilon\to 0$, 
\begin{equation}
J^{\left(\boldsymbol{\rho}\right)}=U^{(\boldsymbol{\rho})}-\frac{2}{|\boldsymbol{\rho}|(|\boldsymbol{\rho}|-1)}\sum_{\substack{b,c \\ 1\leqslant b<c\leqslant L}}\rho_b\rho_c U^{\left(\boldsymbol{1}_{bc}\right)}
\end{equation}
for any $\boldsymbol{\rho}$ with $|\boldsymbol{\rho}|\geqslant3$. The idea is that we can now define the solutions
\begin{equation}
	\tilde{J}^{\left(\boldsymbol{\rho}\right)}=\lim_{\epsilon\to 0}
	J^{\left(\boldsymbol{\rho}\right)}/\epsilon\,.
\end{equation}
In particular, if $L=3$, we obtain that 
\begin{equation}
	\frac{\pi{J}^{\left(111\right)}}{\sin\pi\e}= -\frac{2}{3} \left[V^{[011]}+V^{[101]}+V^{[110]}\right]
\end{equation}
has a finite limit $\e\to0$, while each term in the square brackets contains divergences in components with $|\boldsymbol{a}|=2$. Unfortunately, at $L>3$ the $2^L-L(L+1)/2-1$ solutions $J^{\left(\boldsymbol{\rho}\right)}$ appear to be not linearly independent and, in order to proceed to the full basis of solutions, we have to determine on the next step the linear combinations of $J^{\left(\boldsymbol{\rho}\right)}/\epsilon$ which vanish in the limit $\epsilon\to 0$. The combinatorics appears to be quite involved and we reserve the full consideration for our future work.

\subsection{Equal mass sunrise integral in $D=2$} 

The case of equal mass $L$-loop sunrise integral in $D=2$ has probably received the most attention in the literature, see, in particular, Refs.  \cite{Muller-StachWeinzierlZayadeh2012,Vanhove:2014wqa,BlochKerrVanhove2014,Primo2017,Adams2017b,Broedel2018b}. 

So, we consider the two groups of lines, one containing a single ($r_0=1$) line with $m_0=\sqrt{q^2}$, the other consisting of $r_1=L+1$ lines with mass $m_1=m$. It is convenient to put $q^2=1$.

Similar to Eq. \eqref{eq:V[]}, we define the solutions $V^{[\upsilon]}$ ($\upsilon=0,1,\ldots, L+1$) as follows
\begin{equation}
	V^{[\upsilon]}=\left(-\frac{2}{\pi}\sin \pi\e\right)^{\upsilon}\sum_{\rho_1=\upsilon}^{L+1} {L+1-\upsilon \choose \rho_1-\upsilon} V^{(0\rho_1)}\,,
\end{equation}
where $V^{(\boldsymbol{\rho})}=V^{(\rho_0\rho_1)}$ are defined in Eq. \eqref{eq:Vrho1}. Note that when $\upsilon\geqslant 2$ the function $V^{[\upsilon]}$ is the solution of the homogeneous system for the `cut' sunrise integral. This is because it is a linear combination of $V^{(\boldsymbol{\rho})}$ with $|\boldsymbol{\rho}|>1$ which are all the homogeneous solutions, see footnote \ref{ft:Yrho} on page  \pageref{ft:Yrho}. Again, $00$ components of  $V^{[\upsilon]}$ all have a finite limit at $D=2$:
\begin{gather}\label{eq:Vk}
V^{[\upsilon]}_{00}\stackrel{\epsilon\to 0}{\longrightarrow}2^{L+2}m^{L+1} \intop_0^{\infty} dx\,x
K_{0}\left(x\right)[K_{0}\left(mx\right)]^{L+1-\upsilon}[I_{0}\left(mx\right)]^{\upsilon}\,.
\end{gather}
As we explained in the previous section, the limit of components with $\boldsymbol{a}\neq 00$ of each $V^{[\upsilon]}$ can contain divergences. However, we have some freedom in definition which might help us to get rid of these divergences. Namely, we can construct 
\begin{equation}
	\widetilde{V}^{[\upsilon]}= {V}^{[\upsilon]}+\sum_{\omega=1}^{\upsilon-2} C_{\upsilon \omega}(\e) {V}^{[\upsilon-\omega]}\,,
\end{equation}
where $C_{\upsilon \omega}$ are some coefficients with the asymptotics $ C_{\upsilon \omega}(\e)= O(\e^\omega)$. These new functions $\widetilde{V}^{[\upsilon]}$ have the same properties as  ${V}^{[\upsilon]}$. Namely, they have the same limit of $00$ component as ${V}^{[\upsilon]}$ and are the solutions of the homogeneous system for the cut sunrise graph. Now the idea is that we can fix the coefficients $C_{\upsilon \omega}(\e)$ in such a way as to eliminate all divergences in the limit $\e\to0$. We were able to check this for a few low-loop cases\footnote{Note that this is not the case for the sunrise integral with different masses.}.




Instead of running into complications connected with the basis $V^{(0\rho_1)}$, we can pass to the basis of momenta spanned by the functions $\IKM[\{0,1\}_1,\{\upsilon,L+1-\upsilon\}_m,s]$ defined as
\begin{equation}\label{eq:IKM}
	\IKM[\{a_0,b_0\}_{m_0},\{a_1,b_1\}_{m_1},\ldots,s]=\int dx\, x^s \prod_k [I_0(m_kx)]^{a_k}[K_0(m_kx)]^{b_k}\,.
\end{equation}
Similar functions appear in Refs. \cite{Broadhurst:2016myo,Zhou:2017jnm,Broadhurst:2018tey}.
For any specified $L$ it is easy to construct the matrix $\mathcal{M}$ for transition to moments basis using the matrix $R$ which shifts the power of $x$ in the integrand. Namely, the $n$-th row of the matrix $\mathcal{M}$ has the form
\begin{equation}
\mathcal{M}_n=\left[\prod_{k=1}^{n} R(2+n-k,1)\right]_0
\end{equation}
where $[\ldots]_0$ denotes the zeroth row of the matrix in the brackets. In Appendix \ref{sec:qrels} we present a few examples of the quadratic relations obtained in this way. These quadratic relations resemble those discovered by Broadhurst and Roberts in Ref. \cite{Broadhurst:2018tey}. The difference is that our relations explicitly depend on one parameter $q^2/m^2$ while the relations discovered in Ref. \cite{Broadhurst:2018tey} are obtained for the symmetric point $q^2=m^2$. However, it is not possible to obtain the latter from the former by simply putting $q^2=m^2$. One obstacle is that in the symmetric point (when $q^2=m^2$) the number of independent moments drops by about a half. Another related problem is that our identities tend to develop singularities at $q^2=m^2$. 

Alternatively, we can start from Eq. \eqref{eq:qrelS} for $K=0$, use the shift operators $R$, Eq. \eqref{eq:Dshift}, and $\mathcal{R}_k$, Eq. \eqref{eq:calRk},  and the matrix $\mathcal{M}$ above. In this way we've been able to reproduce the matrix $D_N$ in the left-hand side of Eq. (5.1) of Ref. \cite{Broadhurst:2018tey} up to at least $10$ loops. However, in order to obtain the right-hand side of the same equation, we need the generalization of \eqref{eq:qrelrhs} to the case $r_0>1$.

\section{Conclusion}

In the present paper we considered the $L$-loop watermelon and sunrise integrals regularized both dimensionally and analytically. We have constructed a special set of ``master integrals'' for which we have managed to derive a Pfaffian differential system. 

The basis of solutions of this system is expressed in terms of the Lauricella function $F_C$ with $L$ independent arguments and generic parameters (cf. \cite{Berends:1993ee,Kalmykov:2016lxx}). Thus, we obtain the Pfaffian differential system equivalent to the  Lauricella $F_C^{(L)}$  second-order differential system. To the best of our knowledge, the Pfaffian differential system for $F_C^{(L)}$  was previously known only for $L\leqslant 2$. It worth to note that the obtained system does not have apparent singularities. 
 
Using the symmetry properties of the matrix in the right-hand side of the differential system and following the path similar to that in Ref. \cite{Lee2018}, we obtain the bilinear relations between the solutions of the system for opposite signs of $D$ and $\mub$. 

We derive the operators which shift dimension and/or exponents of the propagators by $\pm 1$. These operators, in particular, allow one to obtain the canonical form of the differential system near any odd integer value of $D$ and integer exponents $\boldsymbol{\alpha}$. Therefore, the expansion of the solutions in the dimensional and analytic regularization parameters in this case can be written in terms of polylogarithmic functions. Thanks to the shift operators, the obtained bilinear relations lead to the quadratic relations between the coefficients of expansion of the solutions near any integer point in $D,\mub$ space.

We have considered the integrals with integer exponents of the denominators. We observe that the quadratic relations for the solutions of the whole differential system do not exist in this case. Instead, we have derived the quadratic relations for the solutions of the homogeneous differential system.

A Pfaffian differential system can be easily restricted to any subvariety in the space of variables. In particular, we have considered the case when the masses and exponents of some lines coincide. In this case the matrix in the right-hand side of the differential system can be naturally rewritten in terms of the generators of $\mathfrak{sl}(2,\mathbb{C})$ acting in the $(r_k+1)$-dimensional irreducible representation ($r_k$ is the number of identical lines in the $k$-th group).

Finally, we have considered the differential system for the tadpole integral in two dimensions ($D=2,\, \alpha_l=1$). The system acquires a block-triangular form, with the nontrivial block of size $2^L-(L+1)L/2$. Using the quadratic relations for the maximally cut sunrise integral with equal masses in $D=2$ we obtain for any given $L$ the quadratic relations for the moments of the product of the Bessel and Macdonald functions. However, we were not able to derive the closed formula for these relations for general $L$. The found relations, in a sense, generalize the quadratic relations obtained by Broadhurst and Roberts \cite{Broadhurst:2018tey} to the case of arbitrary incoming momentum. However, it is not easy to obtain the latter from the former.

The consideration of the present paper, while satisfactorily describing the generic case, can not be considered as exhaustive when it concerns special cases. In particular, we don't fully understand how to construct the basis of solution space and  the quadratic relations for the case $D=2$.
We also did not consider the special configurations of masses when $M$ is not invertible. In particular, it happens in physically relevant case when each group contains even number of identical lines. We underline that once the number of loops $L$ is fixed, one can rely on a brute-force approach to tackle the above issues. The principal problem is to solve them for general $L$.

\acknowledgments
We thank Mikhail Kalmykov for many useful discussions and David Broadhurst for valuable comments on the manuscript. R. Lee would like to express special thanks to the Mainz Institute for Theoretical Physics (MITP) of the Cluster of Excellence PRISMA+ (Project ID 39083149) for its hospitality and support. R. Lee is supported by the grant of the ``Basis'' foundation for theoretical physics.
\appendix
\section{Compatibility conditions}\label{sec:Compatibility}

Let us examine the compatibility conditions of Eqs. \eqref{eq:dlog}, \eqref{eq:alphashift}, and \eqref{eq:Dshift}. First, it is easy to see that $dA=0$, thanks to the identity $[\sigma_l,\sigma_k]=0$. Then we have to check that $[A_l,A_k]=0$, or equivalently, $A\wedge A	= 0$. We have
\begin{multline}\label{eq:AwedgeA}
A\wedge A	={M}^{-1}{dM}W{M}^{-1}\wedge{dM}W+\sum_k\left[\mu_{k}n_{k},{M}^{-1}{dM}\right]W\wedge\frac{dm_{k}}{m_{k}}
\\
={M}^{-1}{dM}\wedge\left[W,dM\right]M^{-1}W-M^{-1}dM\sum_k\left[\mu_{k}n_{k},M\right]M^{-1}W\wedge\frac{dm_{k}}{m_{k}}\\
={M}^{-1}{dM}\wedge\sum_{k}\mu_{k}dm_{k}\left[n_{k},\sigma_{k}\right]M^{-1}W-M^{-1}dM\sum_{k}\mu_{k}\left[n_{k},\sigma_{k}\right]M^{-1}W\wedge dm_{k}=0\,.
\end{multline}
Other conditions have the form
\begin{gather}
	dR_{l}(\mub)+R_{l}(\mub)A(\mub)=A(\mub-2\boldsymbol{e}_{l})R_{l}(\mub)\,,\\
	R_{j}(\mub-2\boldsymbol{e}_{l})R_{l}(\mub)-R_{l}(\mub-2\boldsymbol{e}_{j})R_{j}(\mub)=0\,,\\
	dR(D)+R(D)A(D)=A(D+1)R(D)\,,\\
	R_{l}(D+1,\mub)R(D,\mub)=R(D,\mub-2\boldsymbol{e}_l)R_{l}(D,\mub).
\end{gather}
These identities can be checked along the same lines as Eq. \eqref{eq:AwedgeA}.

Let us also write down two useful identities which allow one to check the symmetry property \eqref{eq:Bsym} of the matrix  $B$ in the right-hand side of the quadratic relation \eqref{eq:qrel1}. They read 
\begin{gather}
	W(D) R(\tilde{D})=R^{\intercal}(D)W(\tilde{D})\,,\\
	W(D,\mub-2\boldsymbol{e}_l)R_l(D,\mub) =-R_l^{\intercal}(-D,-\mub+2\boldsymbol{e}_l)W(D,\mub)\,,\\
	W(D-2,\mub-2\boldsymbol{e})\mathfrak{R}(D,\mub) =(-1)^{L+1} \mathfrak{R}^{\intercal}(2-D,2\boldsymbol{e}-\mub)W(D,\mub)\,.
\end{gather}


\section{Expansion near $D=1$}
\label{sec:D=1}
We will consider $\epsilon$-expansion of $Y$ for $D=1-2\epsilon$. Then the $\epsilon$-expansion of $T$ can be obtained with the help of Eq. \eqref{eq:Yfunctions}. Let us rescale $m_k\to m_k x$ ($k= 1,\ldots, L$) and consider the differential equation with respect to $x$:
\begin{equation}\label{eq:Yequation1}
\frac{dY}{dx}=\epsilon\sum_{a\in \Lambda} \frac{H_{a}}{x-a} Y\,,
\end{equation}
where the set (or alphabet) $\Lambda$ contains $0$ and all letters of the form  
\begin{equation}
	a(\boldsymbol{\eta})=\frac{m_0}{\sum_{k=1}^L \eta_km_k}\,,
\end{equation}
where $L$-tuple $\{\eta_1,\ldots,\eta_L\}$ has elements $\pm1$, and
\begin{align}\label{eq:Hlambda}
H_0&=2\left(N-L-n_0\right)\nonumber\\
H_{a(\boldsymbol{\eta})}&=-2\left(N-1\right)\prod_{k=1}^{L}\frac{1-\eta_k\sigma_0\sigma_k}2
\end{align}
Then we choose the fundamental matrix of solutions of Eq. \eqref{eq:Yequation1} as
\begin{equation}
\tilde{U}(\epsilon,\mb,x)=\lim_{x_0\to0}\mathrm{Pexp}\left[\epsilon\intop_{x_0}^x\sum_{a\in \Lambda} \frac{H_{a}}{y-a}dy\right]x_0^{\epsilon H_0}\,.
\end{equation}
This definition corresponds to the small-$x$ asymptotics 
\begin{equation}
\tilde{U}(\epsilon,\mb,x)\sim x^{\epsilon H_0}\,.
\end{equation}
The $\epsilon$-expansion of the fundamental matrix has the form
\begin{equation}
	\tilde{U}(\epsilon,\mb,x)=\sum_{n=0}^{\infty} \epsilon^n \sum_{a_1\ldots a_n} H_{a_1}\ldots H_{a_n} G(a_1,\ldots,a_n|x)\,,
\end{equation}
where the sum $\sum_{a_1\ldots a_n}$ runs over all words of length $n$ with letters from the alphabet $\Lambda$, and $G(a_1,\ldots,a_n|x)$ denotes the Goncharov's polylogarithms \cite{Goncharov1998,Weinzierl:2002hv} defined recursively by
\begin{align}
	G(a_1,a_2,\ldots,a_n|x)&=\intop_0^x \frac{dy}{y-a_1}G(a_2,\ldots,a_n|y)\,,\\
	G(\underbrace{0,\ldots,0}_n|x)&=\frac{1}{n!}\ln^n x\,.
\end{align}
Let us represent the specific solution $Y_{a_0\boldsymbol{\rho}}$ as 
\begin{equation}
	Y_{a_0\boldsymbol{\rho}}=\tilde{U}(\epsilon,\mb,1) C_{a_0\boldsymbol{\rho}}^{(\varrho_0,\boldsymbol{\rho})}
\end{equation}
where $C_{a_0\boldsymbol{\rho}}^{(\varrho_0,\boldsymbol{\rho})}$ are the integration constants. In order to find $C_{a_0\boldsymbol{\rho}}^{(\varrho_0,\boldsymbol{\rho})}$, we consider the asymptotics $x\to 0$. Using Eqs. \eqref{eq:Vasym1} and \eqref{eq:Yaslim} we have
\begin{multline}
C_{a_0\boldsymbol{\rho}}^{(\varrho_0,\boldsymbol{\rho})}=\lim_{x\to 0}
x^{-\epsilon H_0}Y_{a_0\boldsymbol{\rho}}^{(\varrho_0,\boldsymbol{\rho})}(m_0,x m_k)
=2^{-2\epsilon}\delta_{a_0\varrho_0}\left(-1\right)^{\bar{a}_{0}}
\\
 \times\Gamma\left[1-\left(\bar{a}_{0}-\left|\boldsymbol{\rho}\right|\right)\epsilon\right]	\Gamma\left[\tfrac12-\left(a_{0}-\left|\boldsymbol{\rho}\right|\right)\epsilon\right]
 \prod_{k=1}^{L}(-1)^{\rho_k}\frac{\Gamma\left[\tfrac{1}{2}+(-1)^{\rho_k}\epsilon\right]}{m_0^{2\epsilon\rho_k}m_{k}^{2\epsilon\bar{\rho}_{k}}}\,. 
\end{multline}

\section{Examples of quadratic relations for $\IKM$ functions.}\label{sec:qrels}
Let us present a few examples of the quadratic relations for $\IKM$ functions related to the cut sunrise integral. For $L$-loop sunrise integral the overall number of Bessel and Macdonald functions in the integrand is equal to $L+2$. We use the $\IKM$ functions as defined in Eq. \eqref{eq:IKM}.

For $L=2$ we obtain the following nontrivial quadratic relation:
\begin{multline}
\IKM\left[\{2,1\}_m,\{0,1\}_1,1\right] \IKM\left[\{3,0\}_m,\{0,1\}_1,3\right]\\
-\IKM\left[\{2,1\}_m,\{0,1\}_1,3\right] \IKM\left[\{3,0\}_m,\{0,1\}_1,1\right]\\
=\frac{4 \left(1-5 m^2\right)}{\left(1-m^2\right)^2 \left(1-9 m^2\right)^2}\,.
\end{multline} 
For $L\geqslant3$ we obtain similar relations which are too lengthy to be presented here. These relations, however, simplify on the pseudo-thresholds. 

For $L=3$ we have
\begin{equation}
9 \IKM\left(\{3,1\}_{\frac{1}{4}},\{0,1\}_1,1\right) \IKM\left(\{3,1\}_{\frac{1}{4}},\{0,1\}_1,3\right)-16 \IKM\left(\{3,1\}_{\frac{1}{4}},\{0,1\}_1,1\right){}^2=20\,,
\end{equation}
\begin{multline}
48 \IKM\left(\{2,2\}_{\frac{1}{4}},\{0,1\}_1,1\right){}^2
-27 \IKM\left(\{2,2\}_{\frac{1}{4}},\{0,1\}_1,1\right) \IKM\left(\{2,2\}_{\frac{1}{4}},\{0,1\}_1,3\right)
=20 \pi ^2\,,
\end{multline}
\begin{multline}
-32 \IKM\left(\{2,2\}_{\frac{1}{4}},\{0,1\}_1,1\right) \IKM\left(\{3,1\}_{\frac{1}{4}},\{0,1\}_1,1\right)\\
+9 \IKM\left(\{2,2\}_{\frac{1}{4}},\{0,1\}_1,3\right) \IKM\left(\{3,1\}_{\frac{1}{4}},\{0,1\}_1,1\right)\\
+9 \IKM\left(\{2,2\}_{\frac{1}{4}},\{0,1\}_1,1\right) \IKM\left(\{3,1\}_{\frac{1}{4}},\{0,1\}_1,3\right)=0\,,
\end{multline}
For $L=4$ we have
\begin{multline}
275 \IKM\left(a,\{0,1\}_1,3\right) \IKM\left(b,\{0,1\}_1,1\right)
-275 \IKM\left(a,\{0,1\}_1,1\right) \IKM\left(b,\{0,1\}_1,3\right)\\
-12 \IKM\left(a,\{0,1\}_1,5\right) \IKM\left(b,\{0,1\}_1,1\right)
+12 \IKM\left(a,\{0,1\}_1,1\right) \IKM\left(b,\{0,1\}_1,5\right)\\
=C_1(a|b)
\end{multline}
with 
\begin{gather}
C_1\left(\{3,2\}_{\frac{1}{5}}|\{4,1\}_{\frac{1}{5}}\right)=\frac{5546875}{9216}\,,
\quad
C_1\left(\{2,3\}_{\frac{1}{5}}|\{4,1\}_{\frac{1}{5}}\right)=0\,,\\
C_1\left(\{2,3\}_{\frac{1}{5}}|\{3,2\}_{\frac{1}{5}}\right)=\frac{5546875\pi^2}{18432}\,.
\end{gather}
At $m=1/3$ we have one relation
\begin{multline}
33 \IKM\left(\{2,3\}_{\frac{1}{3}},\{0,1\}_1,3\right) \IKM\left(\{3,2\}_{\frac{1}{3}},\{0,1\}_1,1\right)\\
-33 \IKM\left(\{2,3\}_{\frac{1}{3}},\{0,1\}_1,1\right) \IKM\left(\{3,2\}_{\frac{1}{3}},\{0,1\}_1,3\right)\\
-4 \IKM\left(\{2,3\}_{\frac{1}{3}},\{0,1\}_1,5\right) \IKM\left(\{3,2\}_{\frac{1}{3}},\{0,1\}_1,1\right)\\
+4 \IKM\left(\{2,3\}_{\frac{1}{3}},\{0,1\}_1,1\right) \IKM\left(\{3,2\}_{\frac{1}{3}},\{0,1\}_1,5\right)=\frac{9963 \pi ^2}{2048}\,.
\end{multline}

We note that we could have easily proceeded to more loops.

\bibliographystyle{JHEP}
\bibliography{DEDRRTadpoles}

\providecommand{\href}[2]{#2}\begingroup\raggedright\begin{thebibliography}{10}

\bibitem{Goncharov1998}
A.~B. Goncharov, \emph{Multiple polylogarithms, cyclotomy and modular
  complexes},
  \href{https://doi.org/10.4310/MRL.1998.v5.n4.a7}{\emph{Mathematical Research
  Letters} {\bfseries 5} (1998) 497}.

\bibitem{Weinzierl:2002hv}
S.~Weinzierl, \emph{{Symbolic expansion of transcendental functions}},
  \href{https://doi.org/10.1016/S0010-4655(02)00261-8}{\emph{Comput. Phys.
  Commun.} {\bfseries 145} (2002) 357}
  [\href{https://arxiv.org/abs/math-ph/0201011}{{\ttfamily math-ph/0201011}}].

\bibitem{Adams2017b}
L.~Adams and S.~Weinzierl, \emph{{Feynman integrals and iterated integrals of
  modular forms}},  \href{https://arxiv.org/abs/1704.08895}{{\ttfamily
  1704.08895}}.

\bibitem{Broedel2018c}
J.~Broedel, C.~Duhr, F.~Dulat and L.~Tancredi, \emph{{Elliptic polylogarithms
  and iterated integrals on elliptic curves. Part I: general formalism}},
  \href{https://doi.org/10.1007/JHEP05(2018)093}{\emph{JHEP} {\bfseries 05}
  (2018) 093} [\href{https://arxiv.org/abs/1712.07089}{{\ttfamily
  1712.07089}}].

\bibitem{arXiv0407327v1.math}
H.~A. Verrill, \emph{{Sums of squares of binomial coefficients, with
  applications to Picard-Fuchs equations}},
  \href{https://arxiv.org/abs/0407327v1.math}{{\ttfamily 0407327v1.math}}.

\bibitem{vanhove2019feynman}
P.~Vanhove, \emph{Feynman integrals, toric geometry and mirror symmetry},  in
  \emph{Elliptic Integrals, Elliptic Functions and Modular Forms in Quantum
  Field Theory}, pp.~415--458.
\newblock Springer, 2019.

\bibitem{Berends:1993ee}
F.~A. Berends, M.~Buza, M.~Bohm and R.~Scharf, \emph{{Closed expressions for
  specific massive multiloop selfenergy integrals}},
  \href{https://doi.org/10.1007/BF01411014}{\emph{Z. Phys.} {\bfseries C63}
  (1994) 227}.

\bibitem{Kalmykov:2016lxx}
M.~{\relax Yu}. Kalmykov and B.~A. Kniehl, \emph{{Counting the number of master
  integrals for sunrise diagrams via the Mellin-Barnes representation}},
  \href{https://doi.org/10.1007/JHEP07(2017)031}{\emph{JHEP} {\bfseries 07}
  (2017) 031} [\href{https://arxiv.org/abs/1612.06637}{{\ttfamily
  1612.06637}}].

\bibitem{lauricella1893sulle}
G.~Lauricella, \emph{Sulle funzioni ipergeometriche a piu variabili},
  {\emph{Rendiconti del Circolo Matematico di Palermo (1884-1940)} {\bfseries
  7} (1893) 111}.

\bibitem{kato1988pfaffian}
M.~Kato, \emph{{A Pfaffian system of Appell's F4}}, {\emph{Bull. College Educ.
  Univ. Ryukyus} {\bfseries 33} (1988) 331}.

\bibitem{Goto2013}
Y.~Goto, \emph{{Twisted cycles and twisted period relations for Lauricella's
  hypergeometric function $F_C$}},
  \href{https://doi.org/10.1142/S0129167X13500948}{\emph{International Journal
  of Mathematics} {\bfseries 24} (2013) 1350094}.

\bibitem{Henn2013}
J.~M. Henn, \emph{{Multiloop integrals in dimensional regularization made
  simple}},
  \href{https://doi.org/10.1103/PhysRevLett.110.251601}{\emph{Phys.Rev.Lett.}
  {\bfseries 110} (2013) 251601}
  [\href{https://arxiv.org/abs/1304.1806}{{\ttfamily 1304.1806}}].

\bibitem{Lee2018}
R.~N. Lee, \emph{{Symmetric $\epsilon$- and $(\epsilon+1/2)$-forms and
  quadratic constraints in "elliptic" sectors}},
  \href{https://doi.org/10.1007/JHEP10(2018)176}{\emph{JHEP} {\bfseries 10}
  (2018) 176} [\href{https://arxiv.org/abs/1806.04846}{{\ttfamily
  1806.04846}}].

\bibitem{Adams2013}
L.~Adams, C.~Bogner and S.~Weinzierl, \emph{{The two-loop sunrise graph with
  arbitrary masses}}, \href{https://doi.org/10.1063/1.4804996}{\emph{J. Math.
  Phys.} {\bfseries 54} (2013) 052303}
  [\href{https://arxiv.org/abs/1302.7004}{{\ttfamily 1302.7004}}].

\bibitem{TancrediPC2019}
L.~Tancredi. Private communication.

\bibitem{Muller-StachWeinzierlZayadeh2012}
S.~Muller-Stach, S.~Weinzierl and R.~Zayadeh, \emph{{Picard-Fuchs equations for
  Feynman integrals}},  \href{https://arxiv.org/abs/1212.4389}{{\ttfamily
  1212.4389}}.

\bibitem{Vanhove:2014wqa}
P.~Vanhove, \emph{{The physics and the mixed Hodge structure of Feynman
  integrals}}, \href{https://doi.org/10.1090/pspum/088/01455}{\emph{Proc. Symp.
  Pure Math.} {\bfseries 88} (2014) 161}
  [\href{https://arxiv.org/abs/1401.6438}{{\ttfamily 1401.6438}}].

\bibitem{BlochKerrVanhove2014}
S.~Bloch, M.~Kerr and P.~Vanhove, \emph{A feynman integral via higher normal
  functions},  \href{https://arxiv.org/abs/1406.2664}{{\ttfamily 1406.2664}}.

\bibitem{Primo2017}
A.~Primo and L.~Tancredi, \emph{{Maximal cuts and differential equations for
  Feynman integrals. An application to the three-loop massive banana graph}},
  \href{https://doi.org/10.1016/j.nuclphysb.2017.05.018}{\emph{Nucl. Phys.}
  {\bfseries B921} (2017) 316}
  [\href{https://arxiv.org/abs/1704.05465}{{\ttfamily 1704.05465}}].

\bibitem{Broedel2018b}
J.~Broedel, C.~Duhr, F.~Dulat and L.~Tancredi, \emph{{Elliptic polylogarithms
  and iterated integrals on elliptic curves II: an application to the sunrise
  integral}}, \href{https://doi.org/10.1103/PhysRevD.97.116009}{\emph{Phys.
  Rev.} {\bfseries D97} (2018) 116009}
  [\href{https://arxiv.org/abs/1712.07095}{{\ttfamily 1712.07095}}].

\bibitem{Broadhurst:2016myo}
D.~Broadhurst, \emph{{Feynman integrals, L-series and Kloosterman moments}},
  \href{https://doi.org/10.4310/CNTP.2016.v10.n3.a3}{\emph{Commun. Num. Theor.
  Phys.} {\bfseries 10} (2016) 527}
  [\href{https://arxiv.org/abs/1604.03057}{{\ttfamily 1604.03057}}].

\bibitem{Zhou:2017jnm}
Y.~Zhou, \emph{{Wronskian factorizations and Broadhurst–Mellit determinant
  formulae}}, \href{https://doi.org/10.4310/CNTP.2018.v12.n2.a5}{\emph{Commun.
  Num. Theor. Phys.} {\bfseries 12} (2018) 355}
  [\href{https://arxiv.org/abs/1711.01829}{{\ttfamily 1711.01829}}].

\bibitem{Broadhurst:2018tey}
D.~Broadhurst and D.~P. Roberts, \emph{{Quadratic relations between Feynman
  integrals}}, \href{https://doi.org/10.22323/1.303.0053}{\emph{PoS} {\bfseries
  LL2018} (2018) 053}.

\end{thebibliography}\endgroup
\end{document}